\documentclass[apj]{emulateapj}

\usepackage{graphicx}
\usepackage{amssymb}
\usepackage{aas_macros}
\usepackage{subfig}
\usepackage{natbib}

\shorttitle{The alignment of WASP-32, WASP-38 and WASP-40}
\shortauthors{Brown et al.}

\begin{document}


\title{Analysis of spin-orbit alignment in the WASP-32, WASP-38 and HAT-P-27/WASP-40 systems \footnotemark[$\dagger$]} \footnotetext[$\dagger$]{based on observations (under proposal 087.C-0649) made using the HARPS high resolution {\'e}chelle spectrograph mounted on the ESO 3.6\,m  at the ESO La Silla observatory.}


\author{D. J. A. Brown\altaffilmark{1}}
\email{djab@st-andrews.ac.uk}

\author{A. Collier Cameron\altaffilmark{1}}

\author{R. F. D{\'{\i}}az\altaffilmark{2}}

\author{A. P. Doyle\altaffilmark{3}}

\author{M. Gillon\altaffilmark{4}}

\author{M. Lendl\altaffilmark{5}}

\author{B. Smalley\altaffilmark{3}}

\author{A. H. M. J. Triaud\altaffilmark{5}}

\author{D. R. Anderson\altaffilmark{3}}

\author{B. Enoch\altaffilmark{1}}

\author{C. Hellier\altaffilmark{3}}

\author{P. F. L. Maxted\altaffilmark{3}}

\author{G. R. M. Miller\altaffilmark{1}}

\author{D. Pollacco\altaffilmark{6,7}}

\author{D. Queloz\altaffilmark{5}}

\author{I. Boisse\altaffilmark{8}}

\author{G. H{\'e}brard\altaffilmark{9,10}}


\altaffiltext{1}{SUPA, School of Physics and Astronomy, University of St Andrews, North Haugh, St Andrews, Fife KY16 9SS, UK.}
\altaffiltext{2}{Aix Marseille Universit{\'e}, CNRS, LAM (Laboratoire d'Astrophysique de Marseille) UMR 7326, 13388, Marseille, France}
\altaffiltext{3}{Astrophysics Group, School of Physical \& Geographical Sciences, Lennard-Jones Building, Keele University, Staffordshire, ST5 5BG, UK.}
\altaffiltext{4}{Institut d'Astrophysique et de G{\'e}ophysique, Universit{\'e} de Li{\`e}ge, All{\'e}e du 6 Ao{\^u}t, 17 (B{\^a}t. B5C) Sart Tilman, 4000 Li{\'e}ge, Belgium}
\altaffiltext{5}{Observatoire Astronomique de l'Universit{\'e} de Gen{\`e}ve, 51 Chemin des Maillettes, CH-1290 Sauverny, Switzerland}
\altaffiltext{6}{Astrophysics Research Centre, School of Mathematics \&\ Physics, Queen's University, University Road, Belfast, BT7 1NN, UK.}
\altaffiltext{7}{Department of Physics, University of Warwick, Coventry CV4 7AL}
\altaffiltext{8}{Centro de Astrof{\'i}sica, Universidade do Porto, Rua das Estrelas, 4150-762, Porto, Portugal.}
\altaffiltext{9}{Institut dÕAstrophysique de Paris, UMR7095 CNRS, Universit{\'e} Pierre \& Marie Curie, 98bis boulevard Arago, 75014 Paris, France}
\altaffiltext{10}{Observatoire de Haute Provence, CNRS/OAMP, 04870 St Michel l'Observatoire, France.}


\begin{abstract}
We present measurements of the spin-orbit alignment angle, $\lambda$, for the hot Jupiter systems WASP-32, WASP-38, and HAT-P-27/WASP-40, based on data obtained using the HARPS spectrograph. We analyse the Rossiter-McLaughlin effect for all three systems, and also carry out Doppler tomography for WASP-32 and WASP-38. We find that WASP-32 ($T_{\rm eff}= 6140^{+90}_{-100}$\,K) is aligned, with an alignment angle of $\lambda=10.5^{\circ\,+6.4}_{\,\,\,-6.5}$ obtained through tomography, and that WASP-38 ($T_{\rm eff}=6180^{+40}_{-60}$\,K) is also aligned, with tomographic analysis yielding $\lambda=7.5^{\circ\,+4.7}_{\,\,\,-6.1}$. This latter result provides an order of magnitude improvement in the uncertainty in $\lambda$ compared to the previous analysis of \citet{simpson2011}. We are only able to loosely constrain the angle for HAT-P-27/WASP-40 ($T_{\rm eff}=5190^{+160}_{-170}$\,K) to $\lambda=24.2^{\circ\,+76.0}_{\,\,\,-44.5}$, owing to the poor signal-to-noise of our data. We consider this result a non-detection under a slightly updated version of the alignment test of \citet{brown2012}. We place our results in the context of the full sample of spin-orbit alignment measurements, finding that they provide further support for previously established trends.
\end{abstract}


\keywords{planets and satellites: dynamical evolution and stability
--
stars:individual(WASP-32, WASP-38, WASP-40)
--
techniques: radial velocities
--
techniques: spectroscopic}



\section{Introduction}
\label{sec:intro}
\setcounter{footnote}{10}
The number of planetary systems for which the sky-projected, spin-orbit alignment angle has been measured is steadily increasing, and is now approaching the point at which serious statistical analyses can be made. The majority of these angles have been measured through the Rossiter-McLaughlin (RM) effect \citep{holt1893,schlesinger1910,schlesinger1916,rossiter1924,mclaughlin1924}, a well established technique that considers the small anomaly in the radial velocity curve that is produced by a transit event \citep[e.g][]{queloz2000}. However  there are a growing number of systems for which the misalignment angle has been measured using alternative means. In some cases this is out of necessity, whilst in others it arises from a desire to expand the repertoire of analysis methods that are available, in an effort to reduce the ever-increasing demands on spectroscopic instruments. Examples of the alternatives currently available include analysis of the effect of star spots on the photometric transit observations \citep[e.g.][]{sanchisojeda2011}, consideration of the effect of gravity darkening \citep[e.g.][]{barnes2011}, comparison of the measured and predicted stellar $v\sin I$ \citep[e.g.][]{schlaufman2010}, and Doppler tomography \citep[e.g.][]{cameron2010a}. This last method, whilst not greatly reducing the telescope time required, is able to break the degeneracy between the sky-projected alignment angle and stellar rotation velocity in systems with low impact parameter. It is best suited to analysing hot, rapidly rotating exoplanet host stars, although it can be applied to planetary systems with a range of host parameters. In fact all of the alternative methods are complementary to the RM measurement approach, allowing as they do the study of systems with vastly different properties, and with which that traditional method struggles to cope.

It is becoming increasingly important to push the boundaries of the explored parameter space in this way. The spin-orbit alignment is an excellent diagnostic for competing theories of planetary system formation and exoplanet migration; as the number of systems for which it is measured increases, so too does our understanding of these processes.

The generally accepted scenario has hot Jupiters forming beyond the `snow line' and migrating inwards to their observed separations \citep{sasselov2000}. It is the process by which this migration occurs that is disputed. Loss of angular momentum through interactions with a protoplanetary disk \citep{goldreich1980} was initially proposed as the dominant mechanism, which, if it is assumed that such disks are well-aligned with the stellar spin axis, would produce a population of hot Jupiters in well-aligned orbits. It is worth pointing out, however, that this assumption of aligned protoplanetary disks is increasingly being challenged \citep[e.g][]{bate2010,lai2011,rogers2012} and investigated \citep{watson2011}. The discovery of hot Jupiters in strongly misaligned orbits, including some that are orbiting in a retrograde direction, has also led to the development of competing theories that utilise the Kozai-Lidov mechanism \citep{kozai1962,lidov1962,fabrycky2007,naoz2011}, planet-planet scattering \citep{weidenschilling1996}, tidal friction, or some combination of these processes \citep{fabrycky2007,nagasawa2008,naoz2012}. These mechanisms are naively expected to produce spin-orbit alignment distributions that are closer to isotropic. However the true picture has turned out to be more complex, and appears to lie somewhere between these two extreme distributions.

\citet{winn2010a} found an apparent link between stellar effective temperature and alignment angle; planets in misaligned orbits seem to preferentially orbit `hot' stars ($T_{\rm eff}\geq6250$\,K), whilst aligned planetary orbits seem to be found mostly around `cool' stars ($T_{\rm eff}<6250$\,K). They suggested that this might be connected to the size of the convective envelope, with tidal realignment of orbits around `hot' stars being suppressed owing to their small convective zone. This led \citeauthor{winn2010a} to conclude that the $\lambda$ distribution (at the time of their publication) could be explained by the combination of planet-planet scattering and Kozai-Lidov cycles. 
%However \citet{moutou2011} found that there was no statistically significant difference between the two populations, at least for the limiting temperature chosen by \citet{winn2010a}.

Another potential pattern in the data, and one that was identified early in the development of this sub-field, was that planets with high mass tend to be misaligned (but not retrograde) \citep{johnson2009}. Counter-examples to the initial trend have been found(for example HAT-P-7 \citep{winn2009, narita2009} and WASP-18 \citep{hellier2009,triaud2010}), but planets with $M_p\gtrsim3$\,$M_{\rm Jup}$ do seem to have a different distribution of spin-orbit misalignment angles \citep{hebrard2011}. This was tentatively interpreted as possible evidence for a combination of Kozai-Lidov cycles with tidal circularisation and realignment, but small number statistics were cited as a cautionary factor. More recent work along similar lines has tended to concentrate on the ratio of the planetary mass to the stellar mass rather than the planetary mass in isolation \citep{albrecht2012}.

A more recently discovered correlation is that of alignment angle with host star age. \citet{triaud2011} noticed that, for stars with $M_* \geq1.2M_{\odot}$, all systems older than $2.5$\,Gyr are well-aligned. This implies that the distribution of $\lambda$ changes with time, which in turn suggests that some misalignment mechanism must operate during the youth of hot Jupiter systems, followed by some method of realigning the system as it evolves. If age is the primary factor then having tidal interactions as the governing mechanism for this latter stage would fit with the observed age trend, as planets around older stars will have had longer to tidally realign. On the other hand it may also be that strongly misaligned planets are simply being destroyed much more quickly than their aligned cousins; indeed, such an effect has been theoretically demonstrated for retrograde planets, which are predicted to reach disruption distances several times faster than prograde planets \citep{barker2009,winn2010a}. This would lead to a decrease in the number of hot Jupiters with time, yet \citet{triaud2011} found no such trend. Either tidal realignment occurs faster than orbital decay, or some other mechanism is responsible for the evolution of the distribution of angles that we observe.

\citet{albrecht2012} have re-examined all of these previously detected trends using an updated, homogeneous database of Rossiter-McLaughlin measurements that included their own new measurements and re-analyses. They found that all of the existing trends are consistent with the idea that tidal interactions are responsible for the evolution of the spin-orbit alignment in hot Jupiter systems. They also considered the dependence of $\lambda$ on the scaled orbital distance, finding that it too is consistent with a tide-driven evolutionary picture. Their estimates of characteristic tidal timescales showed that systems which were expected to align rapidly exhibit angles consistent with alignment, whilst those for which tidal realignment was predicted to be weaker display a nearly random distribution of angles. \citeauthor{albrecht2012} stop short however of claiming any mechanism for the production of the initial distribution of $\lambda$ which, from evidence collected so far, seems to be required to be isotropic.

Despite all of this, the question of how hot Jupiters appear where they are and with the spin-orbit angles that they have is far from settled.  There is only so much that `typical' transiting hot Jupiters can tell us; it is the more unusual systems, lying at the extremes of the distributions in mass, effective temperature, and $v\sin I$, and that can only be accessed through methods such as Doppler tomography, that will provide the best test of the theory underlying the evolution of orbital misalignment with time. It is also important to use newer analysis methods to examine systems in tandem with the consideration of the RM effect in order to get to grips with their intricacies, strengths, weaknesses, and inherent error characteristics. In this paper we present new measurements of the spin-orbit alignment angle for three WASP systems. Two, WASP-32 and HAT-P-27/WASP-40 (hereafter WASP-40), have not previously had been analysed. The third system, WASP-38, has been examined before using the RM effect, but we present new spectroscopic data that improves upon the existing parameter uncertainties. For all three systems we compare the results obtained using the RM effect and Doppler tomography. Throughout we will characterise the spin-orbit alignment angle as $\lambda$, following the convention established by \citet{ohta2005} and widely followed in the literature, rather than the alternative convention of $\beta=-\lambda$ used by \citet{triaud2010}.

\begin{table*}
	\begin{center}
	\caption{Existing system parameters for WASP-32, WASP-38, and WASP-40.\label{tab:planets}}
	\begin{tabular}{lllll}
		\tableline \tableline \\
		Parameter    & Unit             & WASP-32                      & WASP-38                & WASP-40 \\ [2pt]
		\tableline \\
		$M_*$           & $M_\odot$ & $1.10\pm0.03$    & $1.203\pm0.036$          & $0.921\pm0.034$ \\ [2pt]
		$R_*$           & $R_\odot$  & $1.11\pm0.05$    & $1.331^{+0.030}_{-0.025}$         & $0.64\pm0.031$ \\ [2pt]
		$T{\rm eff}$  & K                  & $6100\pm100$                 & $6150\pm80$          & $5246\pm153$ \\ [2pt]
		$v\sin I$\tablenotemark{a} & km\,s$^{-1}$  & $5.5\pm0.4$                  & $8.3\pm0.4$            & $2.4\pm0.5$ \\ [2pt]
		$v_{\rm mac}$\tablenotemark{b}  & km\,s$^{-1}$  & $3.5\pm0.3$ & $3.7\pm0.3$ & $1.0\pm0.3$ \\ [2pt]
		$M_p$         & $M{\rm Jup}$  & $3.60\pm0.07$    & $2.691\pm0.058$          & $0.617\pm0.088$ \\ [2pt]
		$R_p$         & $R{\rm Jup,eq}$  & $1.18\pm0.07$    & $1.094^{+0.029}_{-0.028}$ & $1.038^{+0.068}_{-0.050}$ \\ [2pt]
		$P$              & days          & $2.718659\pm0.000008$          & $6.871814\pm0.000045$  & $3.0395589\pm0.0000090$ \\ [2pt]
		$a$              & AU            & $0.0394\pm0.0003$ & $0.07522^{+0.00074}_{-0.00075}$      & $0.03995\pm0.00050$ \\ [2pt]
		$e$              &               & $0.018\pm0.0065$                 & $0.0314^{+0.0046}_{-0.0041}$          & $0$(adopted) \\ [2pt]
		$i$               & deg    & $85.3\pm0.5$      & $88.83^{+0.51}_{-0.55}$         & $85.01^{+0.20}_{-0.26}$ \\ [2pt]
		ref               &                      & $1$                        & $2$, $3$                                     & $4$  \\ [2pt]
		\tableline \\
	\end{tabular}
	\tablenotetext{1}{$v\sin I$ have been updated through spectroscopic analysis of the new HARPS data.}
	\tablenotetext{2}{$v_{\rm mac}$ values were obtained using the \citet{bruntt2010} calibration against $T_{\rm eff}$.}
	\tablerefs{(1) \citealt{maxted2010}; (2) \citealt{barros2011}; (3) \citealt{simpson2011}; (4) \citealt{anderson2011}}
	\end{center}
\end{table*}

\section{Data Analysis}
\label{sec:data}
\subsection{Rossiter-McLaughlin measurements}
\label{sec:RManalysis}
Our analysis method is based around an adapted version of the code described in \citet{cameron2007}. It utilises a Markov Chain Monte Carlo (MCMC) algorithm, and has previously been detailed in \citet{brown2012}. We model the complete sets of photometric and spectroscopic data to maintain consistency, account for parameter correlations, and fully characterise the uncertainties in our results. Since the publication of \citeauthor{brown2012} we have made some small updates to the code to increase its functionality. Rather than using a global stellar `jitter', individual radial velocity (RV) data sets are now allocated `jitter' values individually. Similarly, the line dispersions required for the \citet{hirano2011} formulation for modelling the RM effect are now calculated for each separate set of RV data. We have also updated our stellar radius prior to use the calibration of \citet{southworth2011} rather than that of \citet{enoch2010}, as the former uses a greater number of stars and focuses on a mass range that is directly relevant to exoplanetary systems.

As in \citet{brown2012}, we apply four Bayesian priors in all possible combinations in an attempt to fully characterise the systems under consideration. We apply priors on orbital eccentricity, spectroscopic $v\sin I$, long-term RV trend, and stellar radius (using the method of \citet{enoch2010} in conjunction with the updated coefficients from \citet{southworth2011}). To distinguish between the combinations of priors we consider the reduced spectroscopic $\chi^2$, which we refer to as $\chi^2_{\rm red}$. If there is no combination of priors with a significantly lower value of $\chi^2_{\rm red}$ we choose the model with the fewest free parameters. The application of the stellar radius prior we consider on the basis of the statistical parameter S (the stellar radius penalty) \citep{cameron2007},
\begin{equation}
S=-2\ln P(M_*,R_* ) = \frac{(R_*-R_0)^2}{\sigma_R^2},
\label{eq:S}
\end{equation}
where $M_*$ and $R_*$ are the stellar mass and radius as calculated by the MCMC algorithm, $R_0$ is the stellar radius derived from the (J-H) colour, and $\sigma_R$ is the $1\sigma$ error in $R_0$. S measures the discrepancy between the two stellar radius values, and if we find a large increase in S when the stellar radius prior is removed, we choose a solution in which it is applied as our preferred one.

RV measurements for our new HARPS data were calculated through a Gaussian fit to the cross-correlation functions (CCFs), using a window of three times the FWHM. HARPS spectra cover the wavelength range $378{\rm nm}\leq\lambda\leq691{\rm nm}$.

\subsection{Doppler tomography}
\label{sec:DTanalysis}
Our Doppler tomography method also uses the complete set of photometric and spectroscopic data for an exoplanet system, and is again based around a modified version of the MCMC code discussed by \citet{cameron2007}. In this case however, the alignment of the system is analysed through a comparison of the in-transit CCFs (covering the same $378{\rm nm}\leq\lambda\leq691{\rm nm}$ range as those used for the RV calculations) with a model of the average stellar line profile. This latter model is created by the convolution of a limb-darkened stellar rotation profile, a Gaussian representing the local intrinsic line profile, and a term corresponding to the effect on the line profile of the `shadow' created as the planet transits its host star. This `bump' is time-variable, and moves through the stellar line profile as the planet moves from transit ingress to transit egress. The precise trajectory of the bump is dictated by the impact parameter, $b$, and spin-orbit alignment angle, $\lambda$, which together determine the precise value for the stellar radial velocity beneath the planetary `shadow' at any moment during the transit. This leads to a more accurate model of the spectroscopic transit signature than provided by RM analysis.

As noted previously this provides a powerful method of analysing spin-orbit alignment that is able to explore parameter space unaccessible to the Rossiter-McLaughlin method (such as rapidly rotating host stars) whilst breaking degeneracies inherent in the other method. It has already been used to provide new constraints on the spin-orbit angles of the WASP-3 \citep{miller2010} and HD189733 \citep{cameron2010a} systems, to analyse the WASP-33 \citep{cameron2010b} and CoRoT-11\,b \citet{gandolfi2012} systems, and to examine the alignment in five further systems (WASP-16, -17, -18, -23 and -31)  (Miller et al., \textit{accepted}). 

\section{WASP-32}
\label{sec:W32res}
WASP-32b is a dense hot Jupiter in a $2.72$\,day orbit around a Sun-like (spectral type G, $T_{\rm eff}= 6140^{+90}_{-100}$\,K), lithium depleted star, and is one of only a small number of hot Jupiters with a mass greater than $3$\,Jupiter\,masses. Its discovery was presented by \citet{maxted2010}, who used photometry from WASP-S \citep{pollacco2006} and Faulkes Telescope North (FTN), in concert with spectroscopic observations from the CORALIE spectrograph \citep{queloz2000,pepe2002}, to determine the existence of the transiting planet.

We used the HARPS spectrograph to observe the transit of WASP-32\,b on the night of  2011 September 26. Thirty observations were acquired over the duration of the night, and additional data were collected on the nights of 2011 September 24, 25 and 27 (see journal of observations, Table\,\ref{tab:W32harps}). We obtained simultaneous photometry of the same transit using EulerCam, mounted on the 1.2\,m Leonard Euler telescope at La Silla \citep{lendl2012}, and using the TRAPPIST telescope at La Silla \citep{jehin2011}. We also obtained photometry of a further transit using TRAPPIST, on 2011 November 24.
%The data for these observations can be found in the journal of observations, Tables \,\ref{tab:W32euler} and \ref{tab:W32trappist}.

We carried out a spectroscopic analysis of the new HARPS spectra to determine an updated estimate of $v\sin I$ for the host star. We assumed a macroturbulence of $v_{\rm mac} = 3.5\pm0.3$\,km\,s$^{-1}$ using the calibration of \citet{bruntt2010}. They describe an analytical polynomial correlation between $T_{\rm eff}$ and $v_{\rm mac}$ (their equation (9)) by convolving synthetic line profiles with different $v\sin I$ and $v_{\rm mac}$ values, and fitting to high signal-to-noise spectra from several instruments, including HARPS. We obtained $v\sin I = 5.5\pm0.4$\,km\,s$^{-1}$, in agreement with the value of $4.8\pm0.8$\,km\,s$^{-1}$ found by \citet{maxted2010} from their CORALIE spectra. Our new value was applied as the spectroscopic prior.

\subsection{Rossiter-McLaughlin analysis}
\label{sec:W32_RM}
We initially applied a stellar `jitter' of $1.0$\,m\,s$^{-1}$ to both the existing CORALIE RV data and our new HARPS data; this value is below the level of precision of the spectrographs used for this work, and was added in quadrature to any data points falling within the transit. The values of $\chi^2_{\rm red}$ that were returned by our algorithm with this level of `jitter' applied all fall with $1\sigma$ of $1.0$, indicating that the solution is well constrained. There was therefore no need to increase the level of stellar activity accounted for by our modelling. 

Adding a long-term, linear RV trend produced no discernible effect on the quality of fit that we obtained, or on the value of $\chi^2_{\rm red}$. Relaxing the prior on the stellar radius led to only marginal changes in the values of S, $M_*$, $R_*$, $\rho_*$, and $b$. It also produced no change in the value of $\chi^2_{\rm red}$; we therefore conclude that any such trend is insignificant, and choose as our definitive solution a model which does not apply the prior. Similarly, we elect not to apply the prior on $v\sin I$ in our final solution. Whilst applying the prior produced an increase of $1.0$\,km\,s$^{-1}$ in the value of $v\sin I$ returned by the MCMC algorithm, it had no impact on the value of $\chi^2_{\rm red}$.
 %Any possible trend was found to have a magnitude consistent with $0$\,m\,s$^{-1}$\,yr$^{-1}$ to within $1\sigma$, and we therefore conclude that any such trend is insignificant.
 
The prior on orbital eccentricity required more careful analysis. \citet{maxted2010} reported a marginal $2.8\sigma$ detection of eccentricity in the planet's orbit, and suggested that it could be confirmed through observations of the secondary eclipse. To our knowledge no such observations have been carried out, so we approach the question with all options available. Our models with floating eccentricity all find $e\leq0.014$, slightly less than the value of $e=0.018\pm0.0065$ found by \citet{maxted2010}, and none show any improvement in $\chi^2_{\rm red}$ compared to the equivalent models with fixed, circular orbits. We tested the significance of the eccentricity values recovered by our algorithm using equation\,27 of \citet{lucy1971}, which adopts a null hypothesis of a circular orbit and considers an orbit to be eccentric if this is rejected at the 5\,percent significance level. This F-test indicated that none of the eccentricities are significant, and thus that a circular orbit is favoured.

\begin{figure*}
	\subfloat{
		\includegraphics[width=0.48\textwidth]{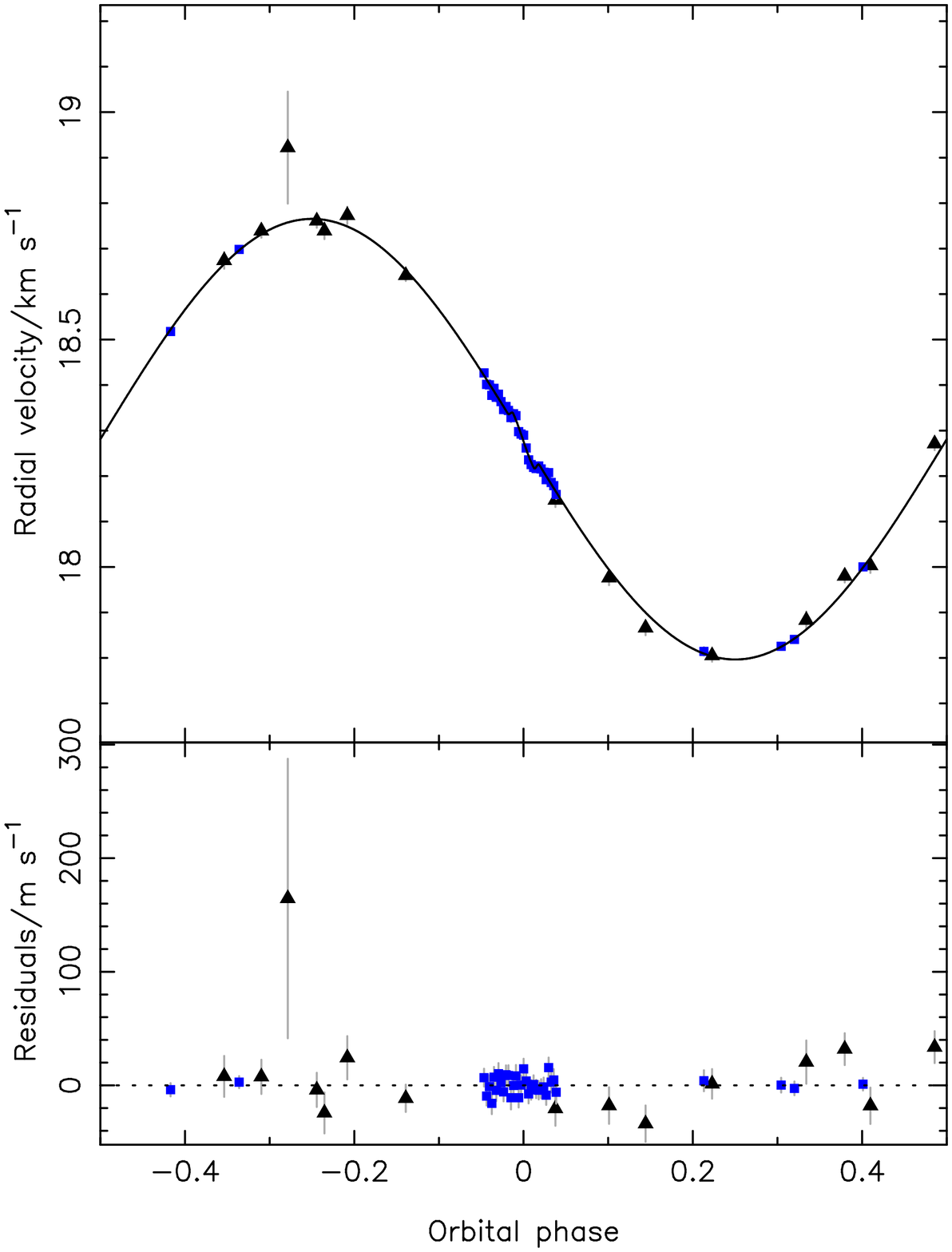}
%		\plotone{fig1a_colour}
%		\plottwo{fig1a.eps}{fig1a_colour.eps}
		\label{fig:W32RV}}
	\subfloat{
		\includegraphics[width=0.48\textwidth]{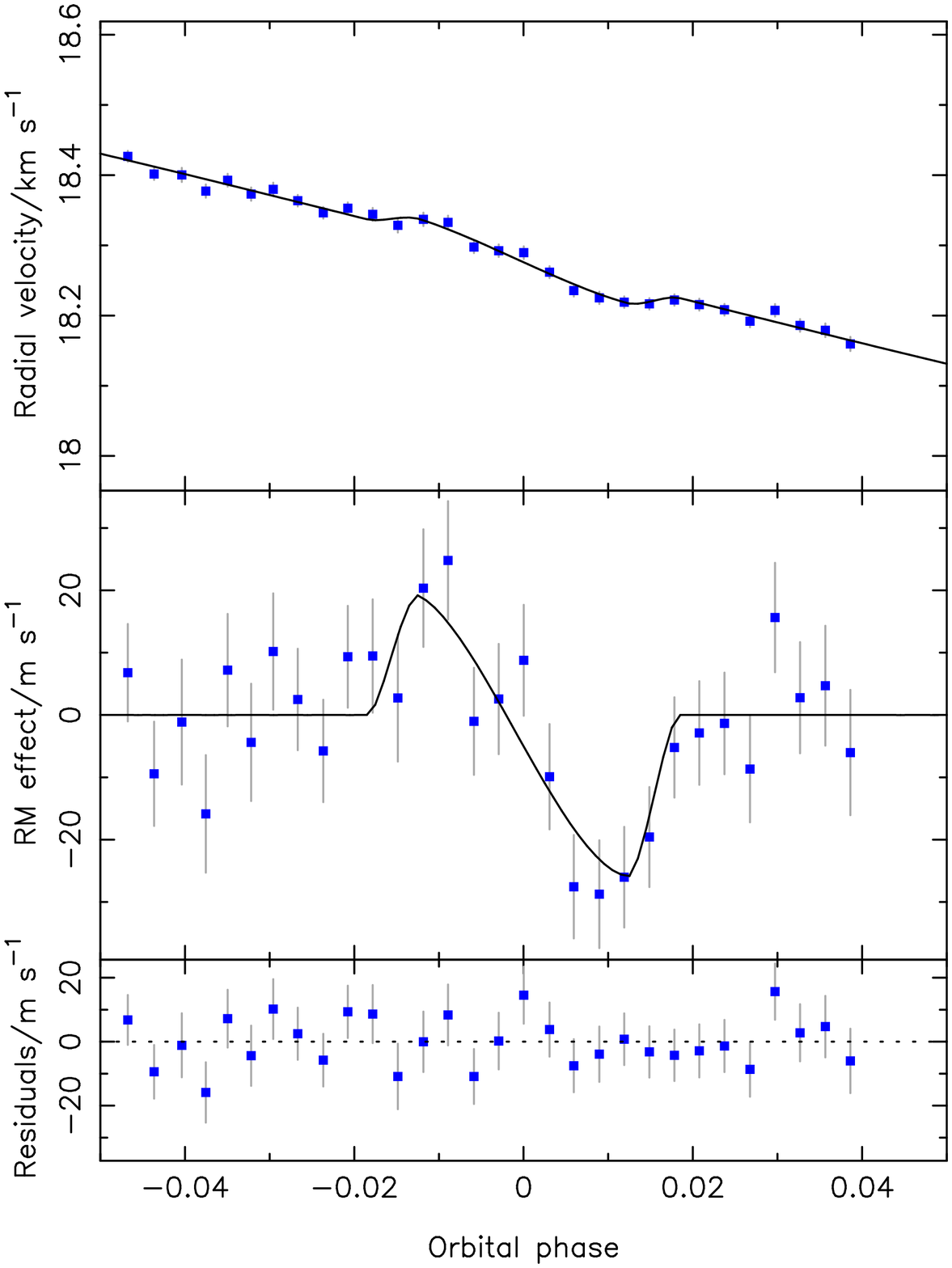}
%		\plotone{fig1b_colour}
%		\plottwo{fig1b.eps}{fig1b_colour.eps}
		\label{fig:W32RM}}
	\caption{Results from our adopted model for WASP-32: $e=0$; no long-term radial velocity trend; no prior on the spectroscopic $v\sin I$, and no stellar radius prior. The best-fitting model is plotted as a solid black line. \textit{Top left:} Complete radial velocity reflex motion curve. Data from CORALIE are denoted by triangles. Data from HARPS are denoted by squares. Error bars are marked in grey; some are smaller than the size of the data points that they accompany. \textit{Bottom left:} Residuals from the RV fit, exhibiting no correlation with phase. \textit{Top right:} Close up of the transit region from the radial velocity curve showing the RM effect, along with the residuals. \textit{Middle right:} Close up of the transit region, with the orbital contribution removed. \textit{Bottom right:} Residuals for the radial velocity data within the RM window. Colour versions of these figures are available in the online edition of the journal.}
	\label{fig:W32res}
\end{figure*}

Our adopted model thus uses the combination of a circular orbit and no long-term RV trend, with neither the $v\sin I$ or stellar radius priors applied. This model provides values of $\lambda=8.6^{\circ\,+6.4}_{\,\,\,-6.5}$, $v\sin I=3.9\pm0.5$\,km\,s$^{-1}$ (slightly slower than the value from spectroscopic analysis), $b=0.66\pm0.02$ and $i=85.1^{\circ}\pm0.2$. The resulting RV curve is displayed in Fig.\,\ref{fig:W32res} alongside a close-up of the transit region, showing the Rossiter-McLaughlin anomaly. The amplitude of the anomaly is low owing to the moderate rotation speed of the host star, but the signal-to-noise is high and the anomaly is well constrained. We found that the semi-amplitudes returned for all three of the RV data sets (the CORALIE data from \citet{maxted2010}, our new HARPS out-of-transit data and our HARPS in-transit data) were in good agreement, and consistent with the results from the discovery paper. Our barycentric velocities on the other hand whilst consistent with each other, are slightly less than the value found by \citet{maxted2010}, even for their CORALIE spectroscopy.

\subsection{Doppler tomography}
\label{sec:W32_DT} 
\begin{figure*}
	\subfloat{
		\includegraphics[width=0.48\textwidth]{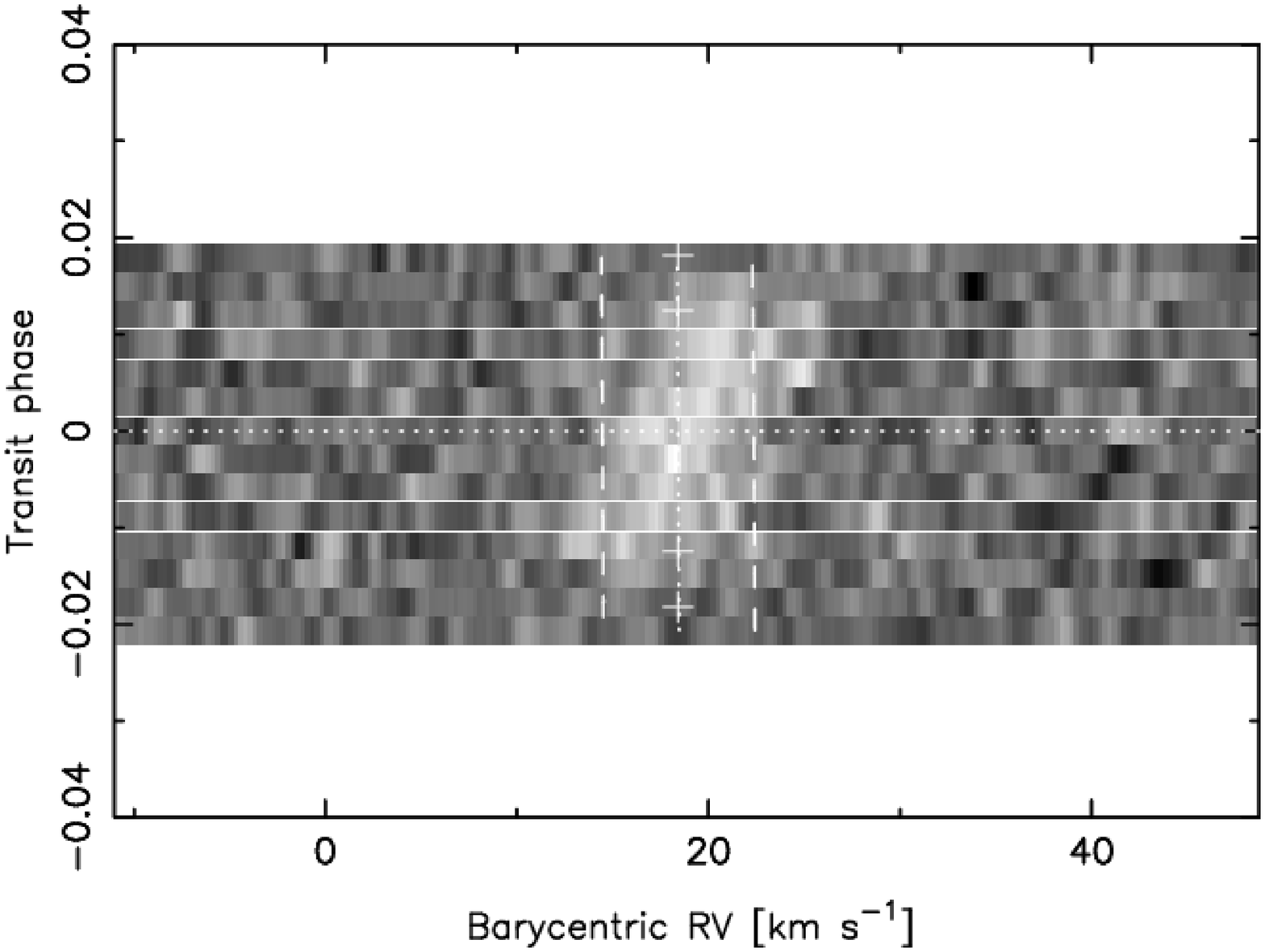}
%		\plotone{fig2a.eps}
		\label{fig:W32bump}}
	\subfloat{
		\includegraphics[width=0.48\textwidth]{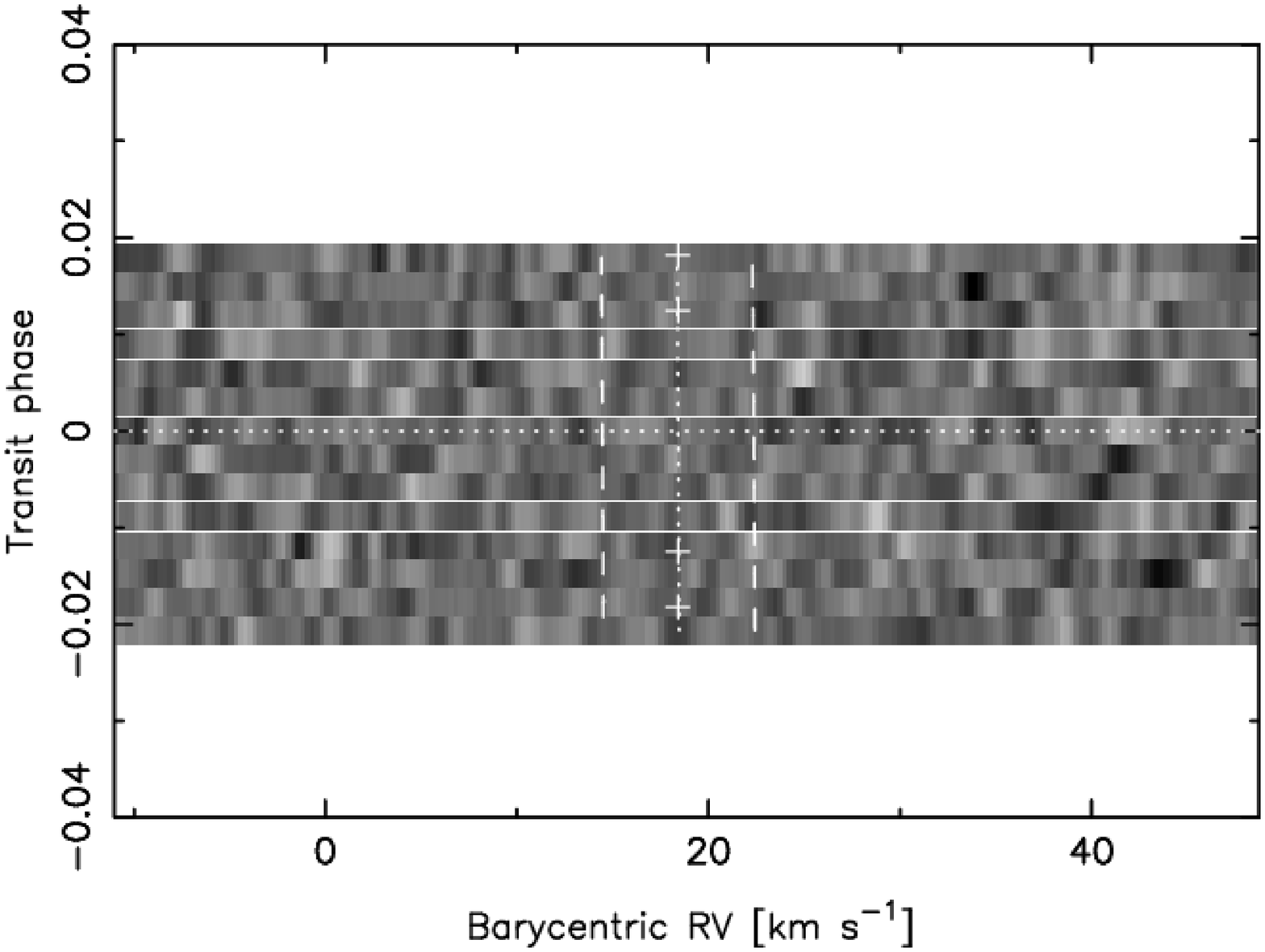}
%		\plotone{fig2b.eps}
		\label{fig:W32resid}}
	\caption{\textit{Left:} Residual map of WASP-32 time series CCFs with the model stellar spectrum subtracted. The signature of the planet moves from bottom-left to top-right, supporting the aligned, prograde orbit conclusion from our RM analysis. \textit{Right:} The best-fitting model for the time-variable planet feature has been subtracted, leaving the overall residual map. The lack of any features in this figure indicate a lack of large-scale stellar activity. \\
	The horizontal dotted line marks the mid-transit phase. The vertical dotted line denotes the stellar radial velocity, whilst the vertical dashed lines indicate $\pm v\sin I$ from this, effectively marking the position of the stellar limbs. The crosses mark the four contact points for the planetary transit.}
	\label{fig:W32dopp}
\end{figure*}

The set of priors identified as comprising the best-fitting model for our RM analysis were applied to our Doppler tomography method, allowing us to assess a single model only. Fig.\,\ref{fig:W32dopp} displays the residual maps from our analysis, which returned values of $v\sin I=3.9^{+0.4}_{-0.5}$\,km\,s$^{-1}$ and $\lambda=10.5^{\circ\,+6.4}_{\,\,\,-5.9}$.

There is little to choose between the results returned by our two analysis methods. In this case, since the constraints on the spin-orbit angle were well-defined by our original RM analysis, the tomography method has been unable to provide much improvement. However it does confirm the results from the traditional, RV measurement based RM analysis, namely that the system is well aligned, and in a prograde orbit. This is easily seen in Fig.\,\ref{fig:W32dopp}. Fig.\,\ref{fig:W32bump} shows the time-series map of the residuals after the subtraction of the stellar line profile only; the effect of the planet therefore shows up as a bright `streak' across the figure, centred on phase 0 and the barycentric radial velocity of the host star, and travelling between the $\pm v\sin I$ values. The trajectory of the planet signature unambiguously identifies the planetary orbit as prograde, moving as it does from bottom-left ($-v\sin I$ at the orbital phase corresponding to ingress) to top-right ($+v\sin I$ at the orbital phase corresponding to egress). Fig.\,\ref{fig:W32resid} in turn displays the final residual map, after the removal of the planet signature. The lack of any notable, consistent deviation from the mean value of the map indicates a lack of significant stellar activity in the host star, as any such activity would produce signatures similar to that of the planet (e.g. non-radial pulsation in WASP-33, \citealt{cameron2010b}). 

\section{WASP-38}
\label{sec:W38res}
The WASP-38 system consists of a massive ($2.7$\,$M_{\rm Jup}$) hot Jupiter in a long ($6.87$\,d), eccentric orbit around a bright (V=9.4), rapidly rotating star of spectral type F8 and $T_{\rm eff}=6180^{+40}_{-60}$\,K. Further information regarding its discovery can be found in \citet{barros2011}. Photometry from the WASP-N array, the RISE instrument mounted on the 2\,m Liverpool Telescope \citep{steele2008,gibson2008} and an 18\,cm Takahashi astrograph at La Palma were combined with spectroscopic measurements taken using the CORALIE and SOPHIE instruments to confirm the presence of the planet.

The Rossiter-McLaughlin effect of WASP-38\,b has been analysed previously by \citet{simpson2011}, who obtained spectroscopic observations of a transit event using the FIES spectrograph mounted on the Nordic Optical Telescope (NOT) at la Palma.  Despite the low precision of their measurements, they were able to place useful constraints on the misalignment angle using the shape of the RV anomaly during transit, ruling out high angles and confining the system to prograde orbits. They reported a final value for the misalignment angle of $\lambda=15^{\circ\,+33}_{\,\,\,-43}$, but were not able to provide a firm conclusions as to the alignment, or otherwise, of the system.

We obtained new spectroscopic observations using HARPS of the transit event on the night of 2011 June 15, as well additional observations made throughout 2011 to provide coverage of the entire radial velocity curve (see journal of observations, Tables\,\ref{tab:W38harps_oot} and \ref{tab:W38harps}). We again obtained photometric observations of a transit using TRAPPIST, on 2011 April 13, but unfortunately we were not able to obtain simultaneous photometry of our spectroscopically observed event. As with WASP-32, we analysed our new HARPS spectra to obtain a value for $v\sin I$ of $8.3\pm0.4$\,km\,s$^{-1}$. A macroturbulence of $v_{\rm mac}=3.7\pm0.3$\,km\,s$^{-1}$ was assumed, again using the calibration of \citeauthor{bruntt2010}. Our $v\sin I$ is in excellent agreement with the values of $v\sin I = 8.6\pm0.4$\,km\,s$^{-1}$ quoted by \citet{barros2011} and $v\sin I = 8.58\pm0.39$\,km\,s$^{-1}$ found by \citet{simpson2011}. Our adopted $v_{\rm mac}$ is significantly lower than the $4.9\pm0.4$\,km\,s$^{-1}$ that \citet{barros2011} used to fit their spectroscopy. \citeauthor{barros2011} used the calibration of \citet{gray2008}, whereas we used that of \citet{bruntt2010}. Reanalysing our new spectra using the \citeauthor{gray2008} calibration returns a slightly lower value of $v\sin I=7.9\pm0.4$\,km\,s$^{-1}$, in agreement with the \citeauthor{barros2011} results. In spite of this, we feel that the \citeauthor{bruntt2010} calibration gives a better fit to our data, and it is therefore that result that we use for our spectroscopic prior.

\subsection{Rossiter-McLaughlin analysis}
\label{sec:W38_RM}
Our initial stellar jitter estimate of $1$\,m\,s$^{-1}$ led to poorly constrained results, with the lowest $\chi^2_{\rm red}$ value returned by any of the models being 1.7. We therefore recalculated the stellar jitter following \citet{wright2005}, obtaining three distinct values. We found that in order to force $\chi^2_{\rm red}\approx 1$ we had to apply the conservative, 20th percentile estimate of $2.1$\,m\,s$^{-1}$ to our new HARPS data, and the 80th percentile estimate of $6.6$\,m\,s$^{-1}$ to the pre-existing FIES, CORALIE, and SOPHIE data\footnote{For an explanation of these estimates, see \citet{wright2005}}.

We found that applying the spectroscopic prior on $v\sin I$ made little difference to the quality of fit that we obtained, or to the values of $v\sin I$ and $\lambda$ that we obtained when compared to the equivalent case without the application of the prior. Similarly, applying a long-term RV trend had no effect on the results, and the magnitude of any possible trend was found to be insignificant at $|\dot{\gamma}|<22$\,m\,s$^{-1}$\,yr$^{-1}$.  The stellar radius prior however, despite producing only a small change in the values of $\chi^2_{\rm red}$, had a significant impact on the results that we obtained. Cases in which the prior was not applied saw average increases in the stellar mass and radius of 6\,percent and 27\,percent respectively over their equivalent cases in which the prior was applied, as well as an average decrease in the stellar density of 49\,percent. Note that these changes do not necessarily match perfectly, as under our model the stellar density is calculated directly from the transit light curves, and independently from the stellar mass and radius. Relaxing the prior also produced significant increases in $v\sin I$, and substantially raised the impact parameter from $\bar{b}=0.15^{+0.33}_{-0.30}$ to $\bar{b}=0.62^{+0.11}_{-0.13}$. Furthermore, we found that removing the prior increased the value of the stellar radius penalty, S, from $\bar{S}=14.0$ to $\bar{S}=105.3$.

Allowing the eccentricity to float led to a clear and significant difference in both $\chi^2_{\rm red}$ and the total $\chi^2$ for the combined photometric and spectroscopic model. The values returned by our algorithm lie at $\geq7\sigma$ from $e=0$, and were found to be significant by the F-test of \citet{lucy1971}. This confirms the eccentricity detection of \citet{barros2011},  and the values that we find are consistent with the value of $e=0.0314^{+0.0046}_{-0.0041}$ reported by those authors.

\begin{figure*}
	\subfloat{
		\includegraphics[width=0.48\textwidth]{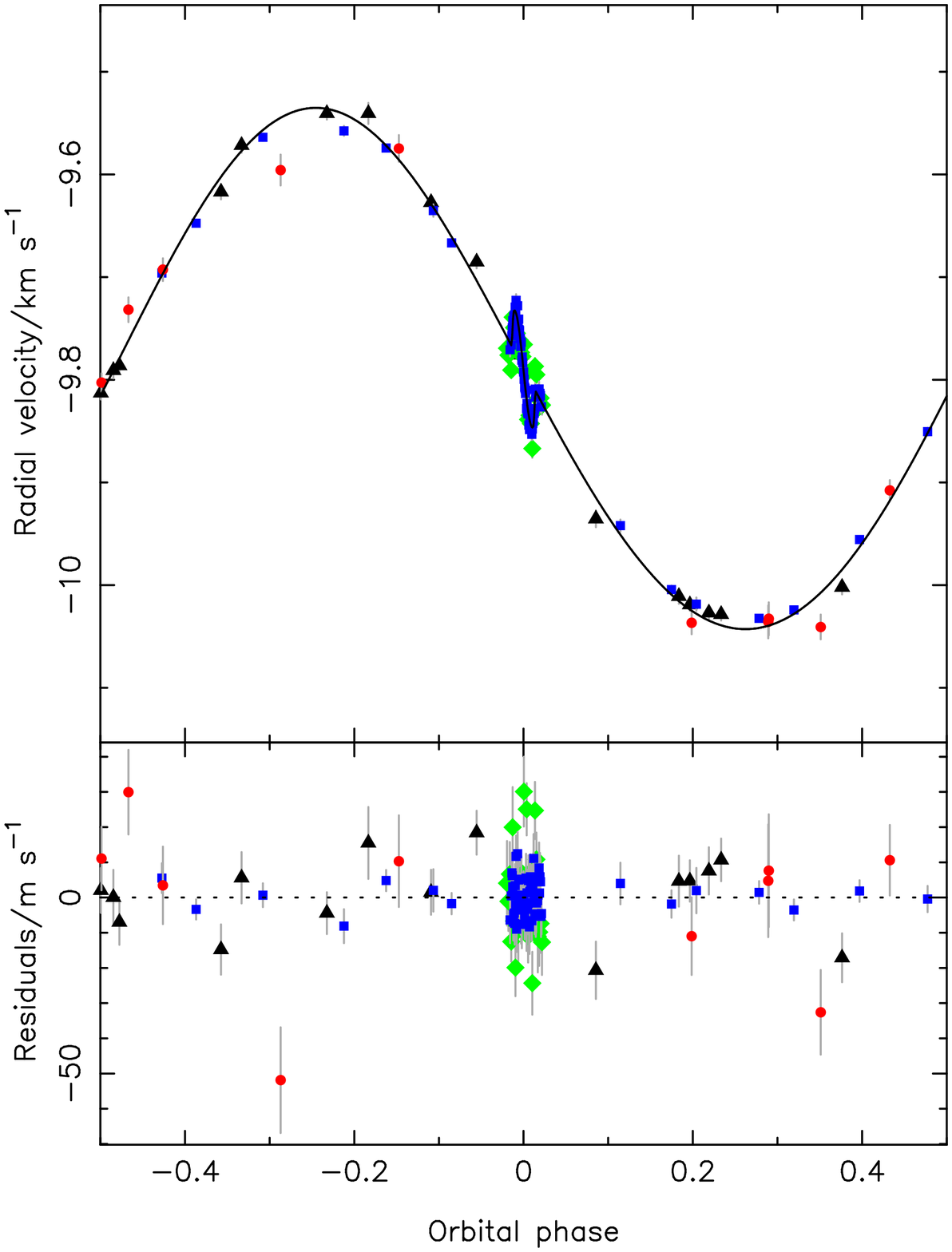}
%		\plottwo{fig3a.eps}{fig3a_colour.eps}
		\label{fig:W38RV}}
	\subfloat{
		\includegraphics[width=0.48\textwidth]{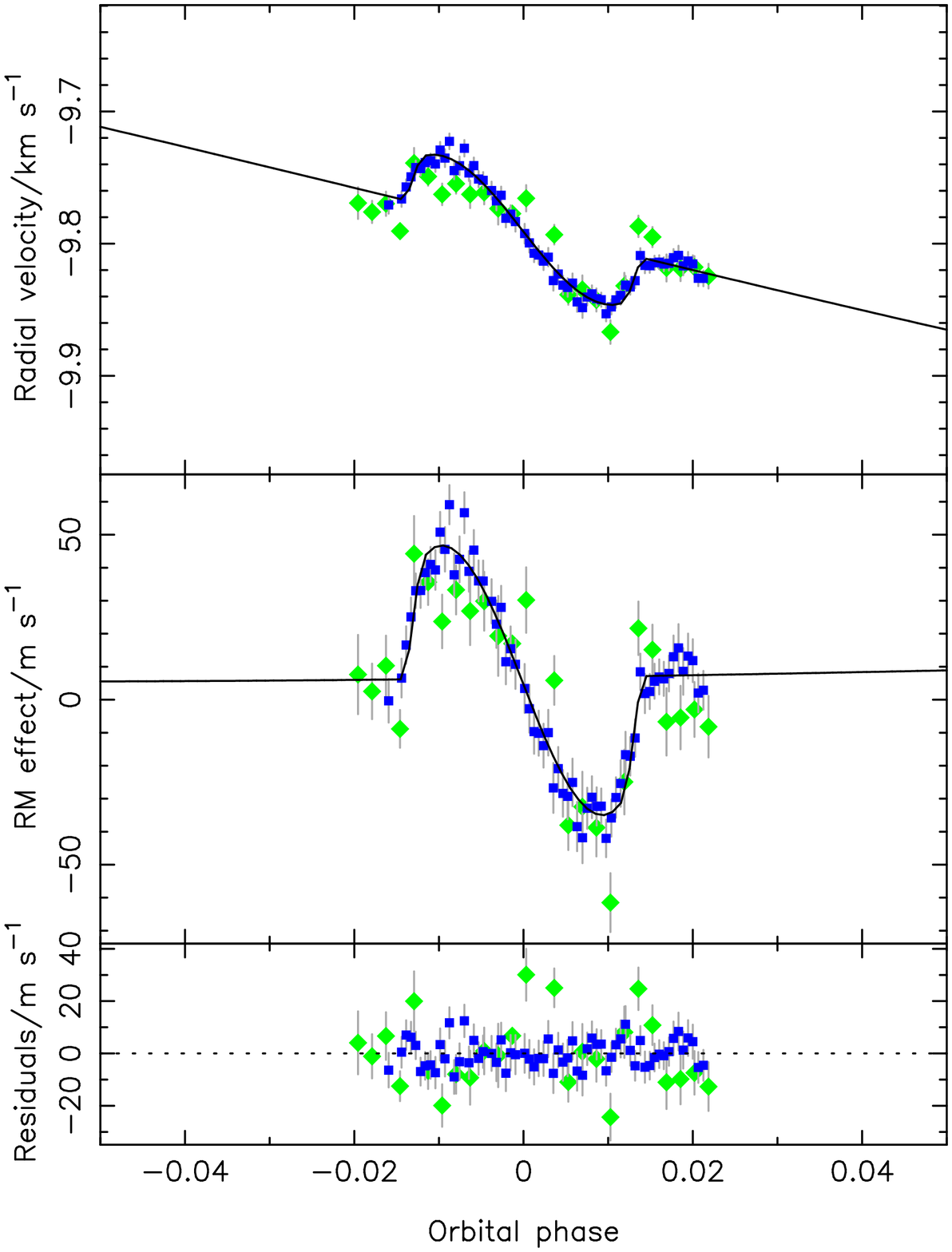}
%		\plottwo{fig3b.eps}{fig3b_colour.eps}
		\label{fig:W38RM}}
	\caption{Radial velocity curve produced by our optimal model for the WASP-38 system. The model uses an eccentric orbit and a prior on the stellar radius, but no long-term radial velocity trend is found and the prior on the spectroscopic $v\sin I$ is not applied. Data from CORALIE are denoted by triangles. Data from SOPHIE are denoted by circles. Data from FIES are denoted by diamonds. Data from HARPS are denoted by squares. Error bars are marked in grey; some are smaller than the size of the data points that they accompany. Format as for Fig.\,\ref{fig:W32res}. Colour versions of these figures are available in the online edition of the journal.}
	\label{fig:W38res}
\end{figure*}

Our adopted model for this system therefore uses an eccentric orbit, does not include a long-term RV trend, does not apply a prior on $v\sin I$, and does utilise a prior on the stellar mass. This model returns values of $\lambda=9.2^{\circ\,+18.1}_{\,\,\,-15.5}$, $v\sin I=7.7^{+0.5}_{-0.4}$\,km\,s$^{-1}$ (slower than the spectroscopic result), $b=0.09^{+0.13}_{-0.06}$ and $i=89.6^{\circ\,+0.3}_{\,\,\,-0.6}$, all of which indicate a well-aligned system. The radial velocity curves are displayed in Fig.\,\ref{fig:W38res}. The difference in quality between the FIES data presented by \citet{simpson2011} and our new HARPS measurements is immediately apparent, particularly during the first half of the transit. The shape of the anomaly is well defined, and it has the large amplitude that is expected given the host star's rapid rotation. It also appears to be highly symmetric, lending credence to the conclusion that the system is likely well-aligned.

We find that the radial velocity semi-amplitudes and barycentric velocities vary somewhat between the 5 different spectroscopic data sets that we analysed (CORALIE data, SOPHIE data, FIES data, HARPS data out-of-transit, and HARPS data in-transit). In particular, the data obtained using FIES by \citet{simpson2011} has a much smaller semi-amplitude than any of the other data sets; $0.152\pm0.030$\,km\,s$^{-1}$ compared to values between $0.246\pm0.001$ and $0.255\pm0.007$\,km\,s$^{-1}$. Interestingly, \citeauthor{simpson2011} found a semi-amplitude of $0.2538\pm0.0035$\,km\,s$^{-1}$ in their analysis, but we suspect that this was overwhelmingly derived from the SOPHIE and CORALIE data, which cover the entire orbital phase. The barycentric velocities agree well with the results of that previous study however.

\subsection{Doppler tomography}
\label{sec:W38_DT}
\begin{figure*}
	\subfloat{
		\includegraphics[width=0.48\textwidth]{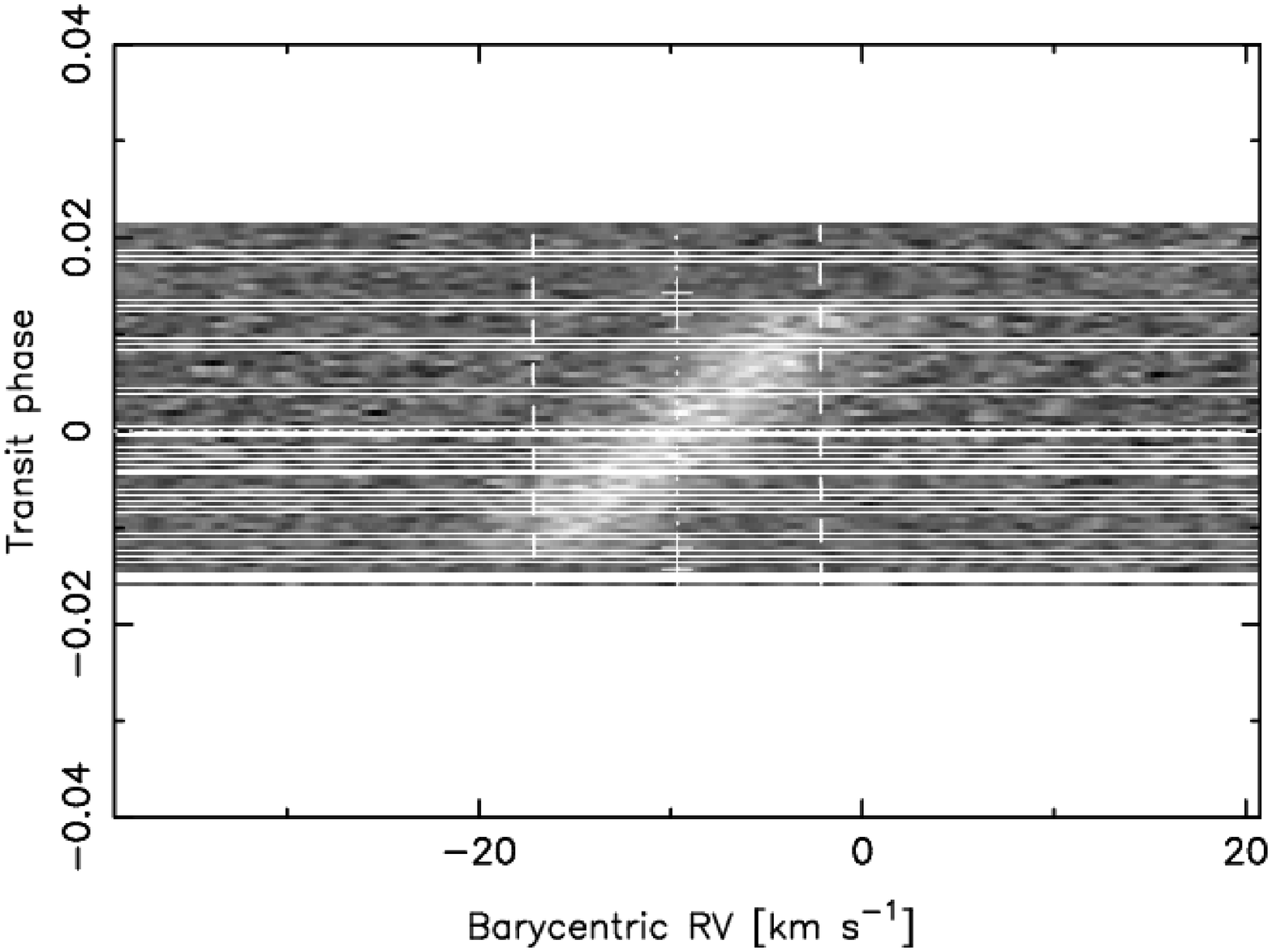}
%		\plotone{fig4a.eps}
		\label{fig:W38bump}}
	\subfloat{
		\includegraphics[width=0.48\textwidth]{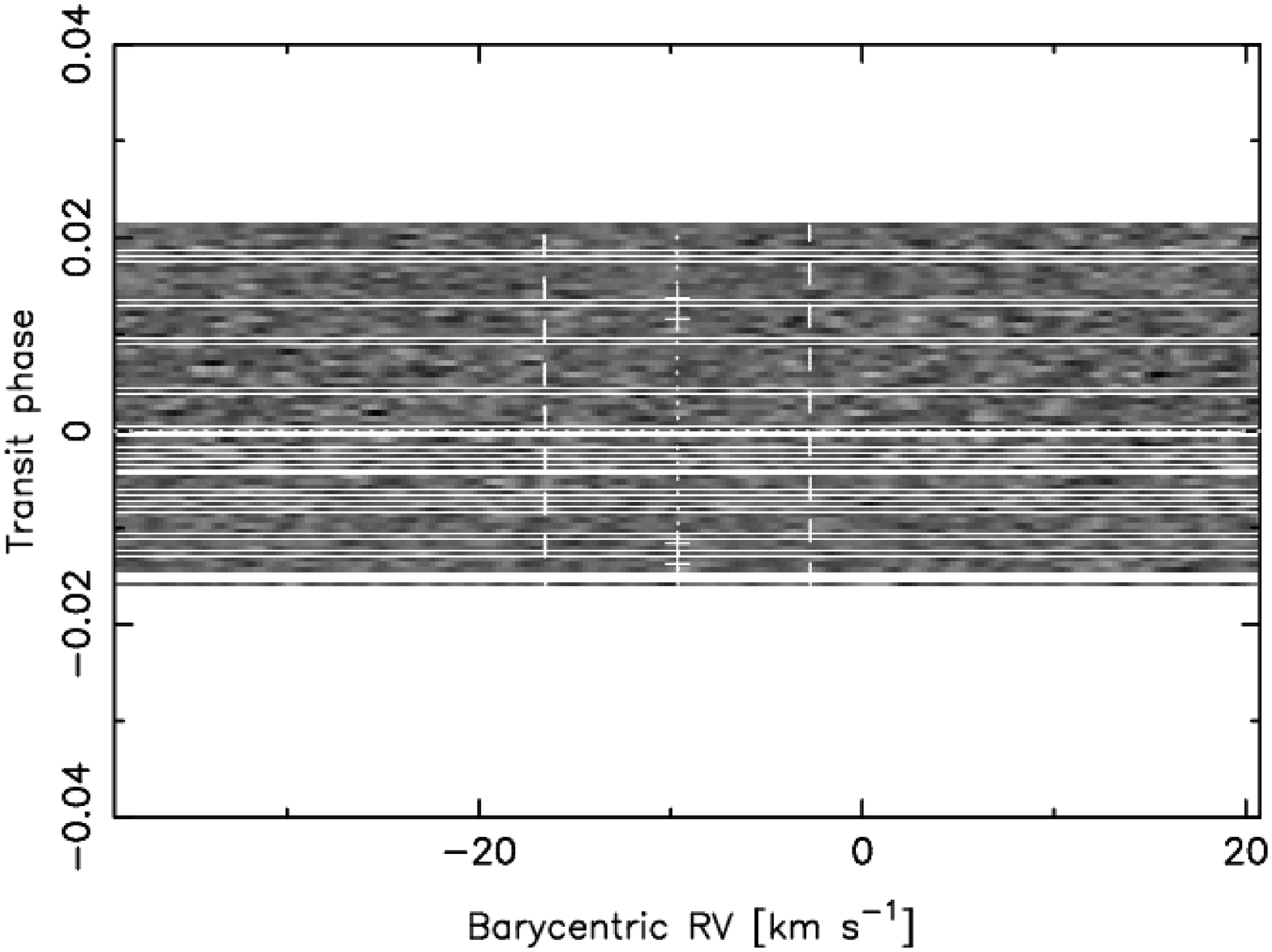}
%		\plotone{fig4b.eps}
		\label{fig:W38resid}}
	\caption{\textit{Left:} Residual map of WASP-38 time series CCFs with the model stellar spectrum subtracted. The bright signature of the planet is clearly visible, and it's trajectory from bottom left to top right clearly indicates a prograde orbit. \textit{Right:} The best-fitting model for the time-variable planet feature has been subtracted, leaving the overall residual map. The lack of any remaining signatures suggests that the star is chromospherically quiet.\\
	Details as for Fig.\,\ref{fig:W32dopp}.}
	\label{fig:W38dopp}
\end{figure*}

We again used the set of priors adopted for our RM modelling as the basis for our Doppler tomography analysis, and Table\,\ref{tab:W38doppres} displays the results from this analysis, together with the results from \citet{simpson2011} and our own RM analysis. It is immediately apparent that we have been able to dramatically reduce the uncertainties on the projected spin-orbit alignment angle; we will return to the question of why this is in Section\,\ref{sec:compare}. The signature of the planet is clearly defined in Fig.\,\ref{fig:W38dopp}, and in the final residual image there is no sign of any anomalies in the stellar line profiles, indicating that the host star is chromospherically quiet.

\begin{table}
%	\caption{Results from our two analysis methods, compared against results from the previous Rossiter-McLaughlin analysis by \citet{simpson2011}. We have significantly reduced the uncertainties on the projected spin-orbit alignment angle compared to their original work.}
	\caption{Comparison of results for WASP-38. \label{tab:W38doppres}}
	\begin{tabular}{llll}
	\tableline \tableline\\
		& $v\sin I$ & $\lambda$   &  $b$\\ [2pt]
	Source & (km\,s$^{-1}$) & (deg) & ($R_*$) \\[2pt]
	\tableline \\
	\citet{simpson2011} & $8.58\pm0.39$            & $15^{+33}_{-43}$        & $0.27^{+0.10}_{-0.14}$ \\ [2pt]
	This work: RM effect & $7.7^{+0.5}_{-0.4}$                 & $9.2^{+18.1}_{-15.5}$ & $0.09^{+0.13}_{-0.06}$ \\ [2pt]
	This work: Tomography & $7.5^{+0.1}_{-0.2}$                 & $7.5^{+4.7}_{-6.1}$   & $0.12^{+0.08}_{-0.07}$ \\ [2pt]
	\tableline \\
	\end{tabular}
\end{table}

\section{HAT-P-27/WASP-40}
\label{sec:W40res}
HAT-P-27 \citep{beky2011} is a fairly typical hot Jupiter system, with a $0.6M_{\rm Jup}$ planet in a $3.04$\,d orbit around a late-G/early-K type star with $T_{\rm eff}=5190^{+160}_{-170}$\,K and super-Solar metallicity. The system was characterised using photometry from HATnet and KeplerCam on the 1.2m FLWO telescope, and spectroscopy from HIRES. It was also independently discovered by the WASP survey using the combined WASP-N and WASP-S arrays, together with spectroscopy from SOPHIE, and designated WASP-40 \citep{anderson2011}.

We obtained new spectroscopic measurements using HARPS of the transit on the night of 2011 May 12, and carried out additional observations at a range of orbital phases throughout 2011 May. New photometric observations were also made using TRAPPIST on 2011 May 17, covering a full transit. We combined these new data with that from both \citet{beky2011} and \citet{anderson2011} for our attempt to characterise the RM effect.

As for WASP-38, we found that our original estimate of $1$\,m\,s$^{-1}$ for the stellar jitter produced poorly constrained ($\chi^2_{\rm red}\approx1.7$) models. We calculated possible values of $5.9$\,m\,s$^{-1}$ (20th percentile), $7.5$\,m\,s$^{-1}$ (median), and $9.8$\,m\,s$^{-1}$ (80th percentile) using the method of \citet{wright2005}, and apply the latter to the existing SOPHIE data. We also note that \citet{beky2011} applied a jitter of $6.3$\,m\,s$^{-1}$ to their HIRES data. To confirm that this was reasonable, we analysed the photometric data in conjunction with only the HIRES spectroscopic data, finding that our initial estimate of $1$\,m\,s$^{-1}$ produced $\chi^2_{\rm red}=10.0\pm1.5$, the 20th percentile value produced $\chi^2_{\rm red}=1.1\pm0.5$, the median value produced $\chi^2_{\rm red}=0.7\pm0.4$, and the 80th percentile value produced $\chi^2_{\rm red}=0.4\pm0.3$, whilst applying their estimate produced $\chi^2_{\rm red}=1.0\pm0.5$. We therefore follow \citeauthor{beky2011} and apply a jitter of $6.3$\,m\,s$^{-1}$ to the HIRES data.

Again, we used our new HARPS spectra to determine $v\sin I=2.4\pm0.5$\,km\,s$^{-1}$, and the calibration of \citet{bruntt2010} to adopt $v_{\rm mac}=1.0\pm0.3$\,km\,s$^{-1}$. This estimate of $v\sin I$ agrees well with the $v\sin I=2.5\pm0.9$\,km\,s$^{-1}$ from \citet{anderson2011}, but is substantially different to the value of $v\sin I=0.4\pm0.4$\,km\,s${-1}$ obtained by \citet{beky2011}, who used $v_{\rm mac}=3.29$\,km\,s${-1}$ based on the calibration of \citet{valenti2005}. We found that using such a high macroturbulence value led to a poor fit for many of the spectral lines, even with $v\sin I=0.0$\,km\,s$^{-1}$, and therefore suggest that \citet{beky2011} have overestimated the broadening in their SOPHIE spectra. The \citet{valenti2005} calibration provides only an upper limit on the macroturbulence, which for cool stars such as WASP-40 can be significantly different from the true values.

As with our RM analysis of WASP-32, we found that there was little to separate the different models for the WASP-40 system, as no significant differences were apparent in the values of $\chi^2_{\rm red}$ that we obtained. The eccentricities returned for models with non-circular orbits were found to be insignificant by the statistical test of \citet{lucy1971}, and the values were all found to be consistent with $e=0$ to within $1.5\sigma$. We also note that the addition of HARPS spectrographic measurements has reduced the value of any possible eccentricity in the system by a factor of 10 compared to the results in \citet{anderson2011}. The addition of a long-term radial velocity trend to the model was found to provide no improvement in the quality of the fit obtained, and the low magnitude of any possible trend ($|\dot{\gamma}|<43$\,m\,s$^{-1}$\,yr$^{-1}$) leads us to conclude that no such trend is present in the system. Imposing a prior on the stellar radius produced only small changes in the mass ($|\delta M_*|\leq2$\,percent), radius ($\delta R_*\leq3$\,percent), and density ($\delta\rho_*\leq7$\,percent) of the host star. The impact parameter was similarly unaffected, with only the error bars increasing with the relaxation of the prior. 

Adding a prior on $v\sin I$ using the spectroscopic measurement produced no change in the value of $\chi^2_{\rm red}$, but it significantly lowered the value of $v\sin I$ returned by the MCMC algorithm, and greatly reduced the uncertainties on the values of $\lambda$ that were being produced. Examination of the HARPS spectroscopy indicated that the amplitude of any RM effect was likely to be low, with the error bars on the data such that they obscured any possible anomaly in the RV curve. This indicated that the value of $v\sin I$ was likely to be low, and that the error bars on $\lambda$ would likely be high. This information, combined with the lack of any difference in the quality of fit, led us to select a solution in which the prior on $v\sin I$ was not applied.

\begin{figure*}
	\subfloat{
		\includegraphics[width=0.48\textwidth]{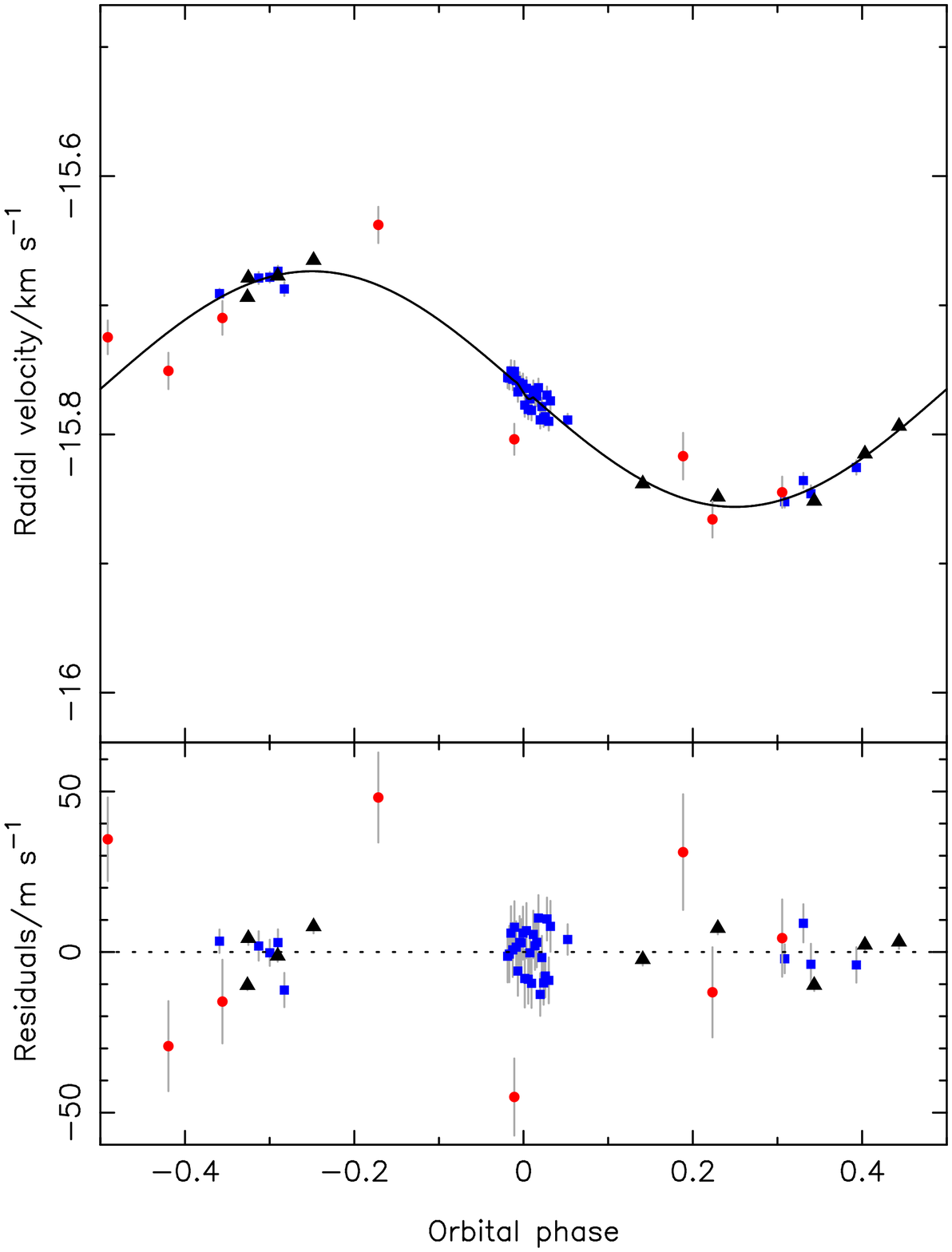}
%		\plottwo{fig5a.eps}{fig5a_colour.eps}
		\label{fig:W40RV}}
	\subfloat{
		\includegraphics[width=0.48\textwidth]{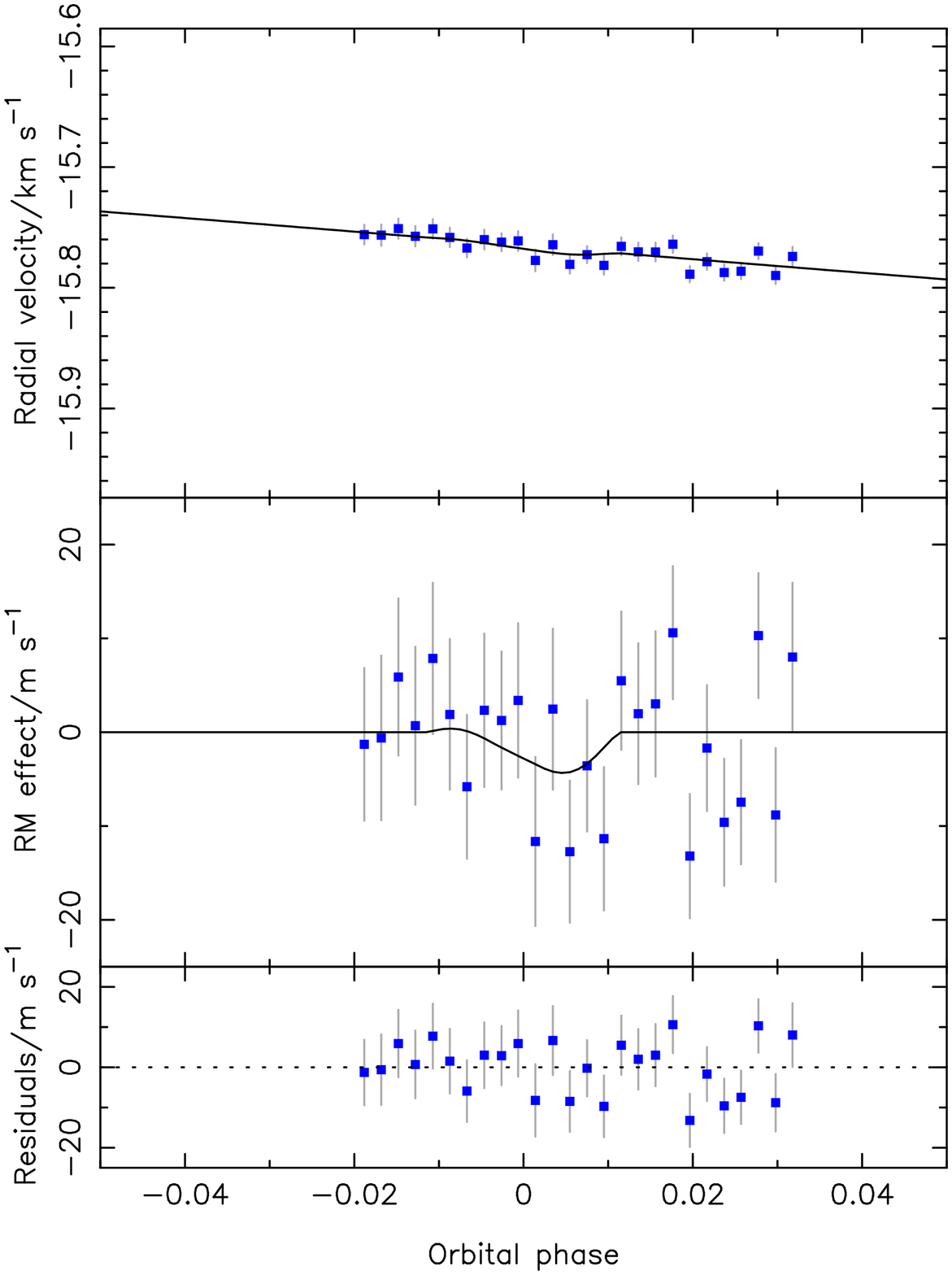}
%		\plottwo{fig5b.eps}{fig5b_colour.eps}
		\label{fig:W40RM}}
	\caption{Results from the fit to the radial velocity data for our adopted solution for WASP-40. A circular orbit was used, with no prior on the spectroscopic $v\sin I$, no long-term radial velocity trend, no prior on the stellar radius. Data from HIRES are denoted by triangles. Data from SOPHIE are denoted by circles. Data from HARPS are denoted by squares. Error bars are marked in grey; some are smaller than the size of the data points that they accompany. Format as for Fig.\,\ref{fig:W32res}. Colour versions of these figures are available in the online edition of the journal.}
	\label{fig:W40res}
\end{figure*}

\begin{figure}
	\includegraphics[width=0.48\textwidth]{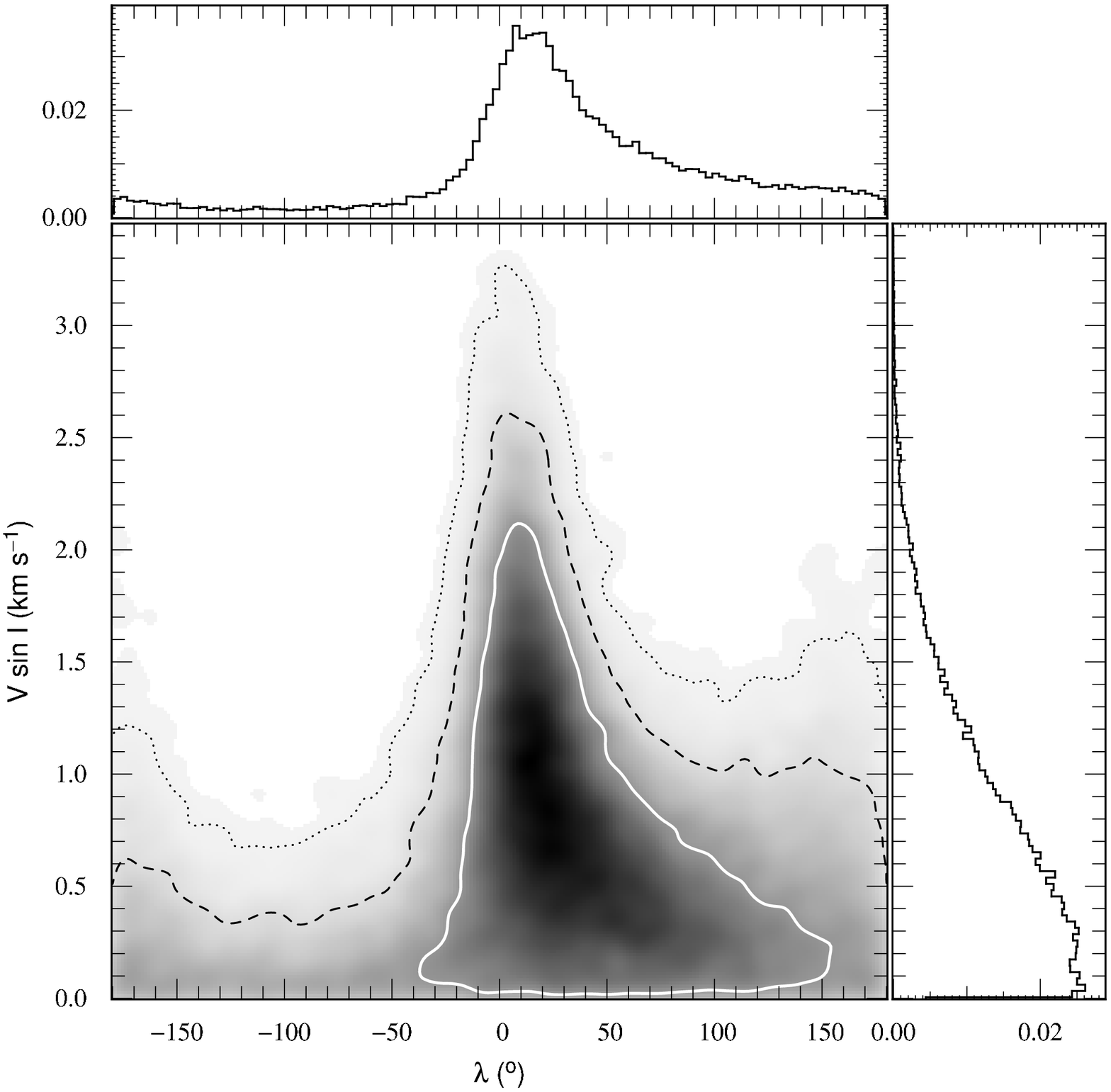}
%	\plotone{fig6.eps}
	\caption{Posterior probability distribution for $v\sin I$ and $\lambda$, derived from the Markov chain, for the fit to the data for WASP-40 described in Fig.\,\ref{fig:W40res}.  The white contour marks the $62.87$\,percent confidence regions, the black, dashed contour the $95.45$\,percent confidence regions, and the black, dotted contour the $99.73$\,percent confidence regions. Marginalised, 1D distributions are displayed in the side panels.  $\lambda=0$ lies well within the main body of the distribution.}
	\label{fig:W40lam-v}
\end{figure}

Our adopted solution therefore uses the combination of a circular orbit and no long-term radial velocity trend, with neither the prior on $v\sin I$ nor the prior on the stellar mass applied. The radial velocity curve that results is shown in Fig.\,\ref{fig:W40res}. Values of $\lambda=24.2^{\circ\,+76.0}_{\,\,\,-44.5}$, $v\sin I=0.6^{+0.7}_{-0.4}$\,km\,s$^{-1}$, $b=0.87\pm0.01$ and $i=85.0^{\circ}\pm0.2$ were returned for this combination of conditions. We note that the value we obtain for the impact parameter is consistent with that found by \citet{anderson2011}, who found a $40.5$\,percent likelihood that the system is grazing. We also note that this system serves as a good example of the systematic discussed by \citet{albrecht2011}. They showed that systems with low-amplitude, low signal-to-noise measurements of the Rossiter-McLaughlin effect were preferentially found to be either close to aligned ($\lambda=0^{\circ}$) or anti-aligned ($\lambda=180^{\circ}$), with the posterior-probability distributions of these systems showing greater ranges of possible solutions around these angles. Fig.\,\ref{fig:W40lam-v} shows the posterior probability distribution for $\lambda$ against $v\sin I$ from our MCMC run. It is immediately clear that there are a greater number of solutions, covering a greater range of values, for $v\sin I$ at angles close to $0^{\circ}$; the effect at $180^{\circ}$ is less pronounced. We note that our solution lies relatively close to the former angle, as predicted by \citet{albrecht2011}, but we also note that our error bars are such that a wide range of alignment angles are included in the possible range of solutions that we find.

In light of this, we analysed the system using our preferred choice of priors and initial conditions, but with no Rossiter-McLaughlin effect fitting. We found that this produced results that showed no difference in terms of quality of fit from our adopted solution, with a value of $\chi^2_{red, noRM}=1.3\pm0.2$ that is in complete agreement with $\chi^2_{\rm red}=1.3\pm0.2$ from the solution adopted above. We therefore consider our weak constraints on the alignment angle to be equivalent to a non-detection of the Rossiter-McLaughlin effect.

\subsection{Doppler tomography}
\label{sec:W40DT}
We attempted to model the system using Doppler tomography, but the combination of the low signal-to-noise and slow rotation proved too difficult to analyse using this method. This nicely highlights a major limitation of the technique, namely systems with poor quality spectroscopic data. RM analysis is able to overcome the poor data quality to provide a result, although it may be inconclusive. However the tomography method is simply unable to process the data if the effect of the planetary transit on the stellar line profile is insignificant.

\section{Our results in context}
\label{sec:discuss}
We now consider our new results, which are summarised in Table\,\ref{tab:summary}, in the context of the complete set of spin-orbit alignment measurements. To date, 52 systems have such measurements published; our results push that number up to 54. 

\begin{table*}
	\caption{Summary of results for WASP-32, WASP-38, and WASP-40 \label{tab:summary}}
	\begin{center}
	\begin{tabular}{lllll}
		\tableline \tableline \\
		Parameter    & Units         & WASP-32\tablenotemark{a} & WASP-38\tablenotemark{a} & WASP-40\tablenotemark{b} \\ [2pt]
		\tableline \\
		\multicolumn{5}{l}{Fitted Parameters} \\  [2pt]
		$D$ & & $0.0108\pm0.0001$ & $0.0069\pm0.0001$ & $0.0143\pm0.0005$ \\  [2pt]
		$K$ & m\,s$^{-1}$ & $0.478\pm0.011$ & $0.252\pm0.004$ & $0.0912\pm0.002$ \\ [2pt]
		$b$ & $R_*$ & $0.66\pm0.02$ & $0.12^{+0.08}_{-0.07}$ & $0.87\pm0.01$ \\ [2pt]
		$W$ & days & $0.0990\pm0.0007$ & $0.1969\pm0.0010$ & $0.070^{+0.001}_{-0.002}$ \\ [2pt]
		$P$ & days & $2.718661\pm0.000002$ & $6.87188\pm0.00001$ & $3.039577^{+0.000005}_{-0.000006}$ \\ [2pt]
		$T_0$ & $BJD_{\rm UTC}-2450000$ & $5681.1945\pm0.0002$ & $5322.1774\pm0.0006$ & $5407.9088\pm0.0002$ \\ [2pt]
		\multicolumn{5}{l}{Derived parameters} \\ [2pt]
		$R_p/R_*$ & & $0.104\pm0.005$ & $0.083\pm0.002$ & $0.120^{+0.009}_{-0.007}$ \\ [2pt]
		$R_*/a$ & & $0.129\pm0.003$ & $0.0829^{+0.0008}_{-0.0007}$ & $0.102^{+0.003}_{-0.004}$ \\ [2pt]
		$R_*$ & $R_{\odot}$ & $1.09\pm0.03$ & $1.35\pm0.02$ & $0.87\pm0.04$ \\ [2pt]
		$M_*$ & $M_{\odot}$ & $1.07\pm0.05$ & $1.23\pm0.04$ & $0.92\pm0.06$ \\ [2pt]
		$\rho_*$ & $\rho_{\odot}$ & $0.84\pm0.05$ & $0.50\pm0.01$ & $1.38^{+0.16}_{-0.13}$ \\ [2pt]
		[Fe/H] & & $-0.13\pm0.10$ & $-0.02\pm0.07$ & $0.14\pm0.11$ \\ [2pt]
		$v\sin I$ & km\,s$^{-1}$ & $3.9^{+0.4}_{-0.5}$ & $7.5^{+0.1}_{-0.2}$ & $0.6^{+0.7}_{-0.4}$ \\ [2pt]
		$R_p$ & $R_{\rm Jup,eq}$ & $1.10\pm0.04$ & $1.09\pm0.02$ & $1.02^{+0.07}_{-0.06}$ \\ [2pt]
		$M_p$ & $M_{\rm Jup}$ & $3.46^{+0.14}_{-0.12}$ & $2.71\pm0.07$ & $0.62\pm0.03$ \\ [2pt]
		$a$ & AU & $0.0390\pm0.0006$ & $0.0758\pm0.0008$ & $0.0400\pm0.0008$ \\ [2pt]
		$i$ & deg & $85.1\pm0.2$ & $89.5^{+0.3}_{-0.4}$ & $85.0\pm0.2$ \\ [2pt]
		$e$ & & $0\mbox{(adopted)}$ & $0.028\pm0.003$ & $0\mbox{(adopted)}$ \\ [2pt]
		$\omega$ & deg & $0$ & $-22.2^{+9.2}_{-8.1}$ & $0$ \\ [2pt]
		$\lambda$ & deg & $10.5^{+6.4}_{-5.9}$ & $7.5^{+4.7}_{-6.1}$ & $24.2^{+76.0}_{-44.5}$ \\ [2pt]
		$|\dot{\gamma}|$ & m\,s\,yr$^{-1}$ & $0\mbox{(adopted)}$ & $0\mbox{(adopted)}$ & $0\mbox{(adopted)}$ \\ [2pt]
		\tableline \\
	\end{tabular}
	\tablenotetext{1}{Results from Doppler Tomography.}
	\tablenotetext{2}{Results from Rossiter-McLaughlin analysis.}
	\end{center}
\end{table*}

\subsection{Effect on existing trends with mass and temperature}
In \citet{brown2012} we updated the $|\lambda|-T_{eff}$ plot of \citet{winn2010a} with all of the systems published since their original analysis, and the results presented in our own work. \citet{albrecht2012} carried out a similar exercise with their new measurements. We consider the new results presented herein in the context of the set of systems listed in the Holt-Rossiter-McLaughlin database compiled by Ren{\'e} Heller\footnote{As of 2012 April 25. http://www.aip.de/People/RHeller}, as well as the new systems and results from \citet{albrecht2012}.

Our new results add little further information to the detected trend with temperature discovered by \citet{winn2010a}. The three systems that we study herein all fit into the `cool' category, although WASP-32 at $T_{\rm eff}=6100\pm100$ lies close to the critical temperature dividing the two sub-populations, and WASP-38 at $T_{eff}=6110\pm150$ encompasses the value of $T_{\rm crit}=6250$\,K within it's $1\sigma$ $T_{\rm eff}$ range.  Similarly, our new results have no effect on the known trend with planetary mass as the systems all have $M_p<3$\,$M_{\rm Jup}$. They therefore cannot provide counter-examples, as planets in this category are already thought to exhibit randomly distributed values of $\lambda$. 

\subsection{Stellar ages}
The host stars of WASP-32 and WASP-40 are insufficiently massive to fulfil the selection criterion imposed by \citet{triaud2011} for his study of the trend of $\lambda$ with stellar age. WASP-38 lies close to the cut-off mass; in some of our simulations it falls below the limit, but in our adopted solution it fulfils \citeauthor{triaud2011}'s criterion for inclusion. We computed the ages for our three systems using a simple isochrone interpolation routine, and several different sets of stellar models, in an attempt to better characterise the inherent uncertainties. Specifically, we made use of the Padova models \citep{marigo2008, girardi2010}, Yonsei-Yale (Y$^2$) models \citep{demarque2004}, Teramo models \citep{pietrinferni2004}, and Victoria-Regina Stellar Structure (VRSS) models \citep{vandenberg2006}. We carried out our isochrone fits in $\rho_*^{-1/3}$ - $T_{\rm eff}$ space, taking the effective temperature from spectroscopic analysis of the HARPS spectra and the stellar density value as found by our preferred model under the tomographic method for WASP-32 and -38, and the RM model for WASP-40. The ages that we obtained for the three systems using these models are set out in Table.\,\ref{tab:ages}. We also assessed the ages of the systems using a combination of the $R'_{HK}$ activity metric, and gyrochronology. We measured the chromospheric Ca II H \& K emission from the new HARPS spectra that we obtained for the three systems discussed herein, using this to calculate $\log(R'_{HK})$. We then computed the stellar rotation period using the method of \citet{watson2010}, which in turn allowed us to estimate the age of the system using the method of \citet{barnes2007}, coupled with the improved coefficients of \citet{meibom2009} and \citet{james2010}.

\begin{table*}
	\caption{Age estimates for the three systems.\label{tab:ages}}
	\begin{center}
	\begin{tabular}{llllll}
		\tableline \tableline \\
						& \multicolumn{4}{c}{Stellar model fitting age} 													& Gyrochronology \\ [2pt]
		                                      & Padova                             & Y$^2$                               & Teramo                             & VRSS                                & age				\\ [2pt]
		System			& (Gyr)				& (Gyr)				& (Gyr)				& (Gyr)				& (Gyr)              			\\[2pt]
		\tableline \\
		WASP-32                    & $2.36^{+1.72}_{-0.85}$ & $2.22^{+0.62}_{-0.73}$ & $4.50^{+1.88}_{-1.69}$ & $1.41^{+1.36}_{-1.10}$ & $2.42^{+0.53}_{-0.56}$ \\ [2pt]
		WASP-38                    & $3.41^{+0.48}_{-0.43}$ & $3.29^{+0.42}_{-0.53}$ & $3.59^{+0.77}_{-0.70}$ & $3.20^{+0.73}_{-0.59}$   & $3.41^{+0.26}_{-0.24}$ \\ [2pt]
		WASP-40                    & $>1.20$                            & $6.36^{+5.86}_{-3.11}$      & $>4.96$                            & $>5.73$                            & $3.60^{+1.78}_{-1.84}$ \\ [2pt]
		\tableline \\
	\end{tabular}
	\end{center}
\end{table*}

The age estimates for the WASP-32 system vary quite considerably but do all agree within the $1\sigma$ errors, with the age from gyrochronology lying in the middle of the range. However this is more a function of the rather large uncertainties than any indication that the age is well constrained. The system nicely highlights the dependence of isochronal age estimates on the set of stellar models that is used. As discussed by \citet{southworth2009}, the choice of stellar models can have a significant impact on the derived properties of exoplanetary systems, particularly through the introduction of systematic errors. \citet{southworth2009} also notes though that different sets of models are often based on the same physical underpinnings, differing only in their implementation, and that therefore the results cannot be considered to be totally independent. \citet{triaud2011} pointed out that isochronal analysis is less precise for stars with $M_*<1.2$\,$M_{\odot}$ owing to the increased length of their main sequence lifetime, and their less pronounced radius increase (and therefore density decrease) than more massive stars. We found a mass of $1.11\pm0.02$\,$M_{\odot}$ from our RM analysis, in agreement with the value from \citet{maxted2010}; the wide age range that we obtain is therefore expected given our preceding comments. The WASP-38 system on the other hand shows much better agreement between the age estimates obtained using the different sets of stellar models. With a mass of $1.17\pm0.02$\,$M_{\odot}$ from our RM analysis, the system lies closer to the arbitrary cut-off of \citet{triaud2011}, so we might expect that the age would be better constrained. Nevertheless, our four age estimates for WASP-38 all agree with the postulated trend for alignment angle to decrease with time

 Finally, WASP-40 is poorly constrained, and we are unable to place upper limits on the age using the available isochrones for 3 out of the 4 model sets that we tried. It is hard to conclude anything from this, but the different models do agree that the system is older than either WASP-32 or WASP-38. On the other hand, we note that the gyrochronological estimate of the stellar age is significantly lower than the age limit that we found from our isochronal fits to the Teramo and VRSS models, although it is consistent with the results from the Y$^2$ and Padova models. \citet{anderson2011} found ages for the system of $6\pm5$\,Gyr using the stellar models of \citet{marigo2008} and \citet{bertelli2008}, which is consistent with our values. They too found a lower age ($1.2^{+1.3}_{-0.8}$\,Gyr) using gyrochronology, in their case based on an estimate of the rotation period derived from $v\sin I$, but this does not match our estimate. From this information we tentatively predict, following the trend noticed by \citet{triaud2011}, that the system will prove to be aligned if the uncertainty on $\lambda$ is able to be reduced. 
 
\subsection{Are the systems aligned?}
\label{sec:aligned}
In \citet{brown2012} we introduced a new test for misalignment that is based upon the Bayesian Information Criterion (BIC). The BIC is calculated for the set of radial velocity measurements that fall within the transit window, for both the best-fitting model and one that assumes an aligned orbit with $\lambda=0^{\circ}$. The ratio, $B$, of the aligned orbit BIC to the free-$\lambda$ BIC is then calculated. Systems with $B\leq0.99$ are classed as aligned, systems with $B\geq1.01$ are classed as misaligned, and systems that fall between these limits are classed as indeterminate. \citet{albrecht2012} note that the BIC test is affected by the relative numbers of RV measurements in transit compared to out of transit, and that it assumes that no correlated noise is present. We acknowledge that these are indeed shortcomings of our test, and that they might affect the boundaries between the three categories discussed in \citet{brown2012}, but we comment that the test is still quantitative, as opposed to the qualitative nature of the previous tests in \citet{triaud2010} and \citet{winn2010a}.

We applied this test to our new results, calculating values of $0.92$ for WASP-32, indicating alignment, $0.95$ for WASP-38, indicating alignment, and $0.97$ for WASP-40, indicating alignment. However in \citet{brown2012} we also postulated a fourth category, that of `no detection', defining this as $v\sin I$ consistent with $0$ to within $1\sigma$. With further reflection, we consider this definition to be inadequate. Our analysis routines are set-up in such a way that such a scenario is highly unlikely to exist; indeed, a lower error bar on $v\sin I$ of greater magnitude than the value itself is nonsensical, as negative rotation is a physical impossibility when considering only the magnitude of the rotation. We therefore revise this category of `no detection' to include systems with $v\sin I$ consistent with $0$ to within $2\sigma$. This new definition encompasses WASP-40, as we feel is appropriate given the poor signal-to-noise of the data that we obtained, and the indistinct Rossiter-McLaughlin effect that we find. The amended category includes no additional systems from our previous sample (see \citet{brown2012} for details), although WASP-1 and WASP-16 are close, but would include the  results for TrES-2 \citep{winn2008} and HAT-P-11 \citep{winn2010b}.

%\begin{figure*}
%	\subfloat{
%		\includegraphics[width=0.48\textwidth]{NotUsed/fig7a.eps}
%		\plottwo{fig7a.eps}{fig7a_colour.eps}
%		\label{fig:BICall}}
%	\subfloat{
%		\includegraphics[width=0.48\textwidth]{NotUsed/fig7b.eps}
%		\plottwo{fig7b.eps}{fig7b_colour.eps}
%		\label{fig:BICzoom}}
%	\caption{BIC ratio, $B$, as a function of $\lambda$ for the full set of planets considered in \citet{brown2012}, as well as the systems presented in this study. WASP-32 is denoted by a filled, blue triangle. WASP-38 is denoted by a filled, red diamond. WASP-40 is denoted by a filled, green square. \textit{Left: } All data. \textit{Right: } A close-up of the heavily populated region in the lower left of the plot, around $B=1.00$ and $|\lambda|=0^{\circ}$. The horizontal dotted lines mark $B=0.99$ and $B=1.01$, the divisions between our alignment categories. The vertical dotted line denotes $|\lambda|=0^{\circ}$. Colour versions of these figures are available in the online edition of the journal.}
%	\label{fig:BIC}
%\end{figure*}

\subsection{Tidal timescales}
\citet{albrecht2012} present two different approaches for estimating the tidal evolution timescales for hot Jupiter systems, and calculate said timescale for a large sample of planets for which the Rossiter-McLaughlin effect has been measured. They take two approaches. In the first, they consider a bimodal sample of planets: those with convective envelopes, and those with radiative envelopes. In the second approach they consider the mass of the convective envelope, which they link to stellar effective temperature. Unfortunately this second approach relies on an unspecified proportionality constant, and the relation between $T_{\rm eff}$ and $M_{\rm CZ}$ that they derived is also unknown. We therefore consider their first approach, which is encapsulated in the equations

\begin{equation}
\frac{1}{\tau_{\rm CE}} = \frac{1}{10\cdot 10^9 {\rm yr}} q^2 \left(\frac{a/R_*}{40}\right)^{-6},
\label{eq:tauCE}
\end{equation}and
\begin{equation}
\frac{1}{\tau_{\rm RA}} = \frac{1}{0.25\cdot 5\cdot 10^9 {\rm yr}} q^2 \left(1+q\right)^{5/6} \left(\frac{a/R_*}{6}\right)^{-17/2}.
\label{eq:tauRA}
\end{equation}

The stellar effective temperatures of our three systems are, as mentioned previously, below the critical temperature dividing the `hot' and `cool' regions of parameter space. They therefore all fall under the convective envelope version of the tidal timescale equation. Using parameters from our best-fitting models (tomographic for WASP-32 and WASP-38, and RM for WSAP-40), we calculate the tidal timescales for our three systems using \ref{eq:tauCE}. We find $\tau_{\rm CE}=5.46867636\times 10^{10}$\,yr for WASP-32, $\tau_{\rm CE}=1.79351519\times 10^{12}$\,yr for WASP-38, and $\tau_{\rm CE}=5.41028372\times 10^{12}$\,yr for WASP-40. These values fit nicely into the scheme that \citet{albrecht2012} developed, whereby systems in which the tidal timescale is short preferentially show low values of $\lambda$, whereas those with longer timescales appear to present an almost random distribution of $\lambda$. The timescales for our three systems are relatively short, particularly where WASP-32 is concerned, and the small alignment angle that we obtained for that system is exactly as expected.

\section{Why use Doppler Tomography?}
\label{sec:compare}
As we discussed in Section\,\ref{sec:intro}, Doppler tomography is one of a number of methods for characterising spin-orbit alignment that are beginning to be used as alternatives to the traditional radial velocity based approach that we have used to analyse all three of the systems in this study. Although tomography has weaknesses, and cannot be applied to every planetary system (as witnessed with WASP-40 previously), it has one great selling point over the radial velocity method. Tomography is able to lift the strong degeneracy that exists between $v\sin I$ and $\lambda$, and which is strongest in systems with low impact parameter.

The geometry of the path that the planetary orbit traces across the stellar disc affects the uncertainty in the spin-orbit alignment angle, particularly if that path is symmetric with respect to the approaching and receding hemispheres of the star. As the impact parameter increases, the range of alignment angles than can produce a symmetric RM curve decreases \citep{albrecht2011}. The limiting cases illustrate this well. With $b=0$, any value of $\lambda$ will produce equal transit path lengths through the red- and blue-shifted halves of the stellar disc, whilst with $b=1$, only $\lambda=0^{\circ}$ and $\lambda=180^{\circ}$ will have the same effect. Thus as $b$ decreases, the uncertainty in the estimate of $\lambda$ increases.

This is not the only parameter involved however. The stellar rotation, $v\sin I$, dictates the amplitude of the Rossiter-McLaughlin anomaly, but this is often ambiguous owing to the uncertainties present in the radial velocity measurements. It is often not clear, particularly for systems with low $v\sin I$, whether the anomaly is truly asymmetric, or whether it is an effect produced by the error bars (see for example WASP-25 in \citet{brown2012}). This means that the same anomaly can often be fit in two different ways. Either $v\sin I$ is low and $\lambda$ indicates misalignment, with the resulting asymmetry in the model used to fit the uncertainties, or $\lambda$ is low and a rapid $v\sin I$ is used, with the greater amplitude providing the required fit. Often what results is a compromise solution, with large error bars on both parameters and some degree of degeneracy between them. This arises owing to the use of the Rossiter-McLaughlin effect to characterise both parameters simultaneously. The problem is exacerbated for systems with low signal-to-noise, for which the range of possible models that fit the data is greatly increased owing to the greater relative size of the uncertainties, and for systems with low impact parameter, for the reasons discussed above. 
 
The Doppler tomography method does not suffer from this same problem, and is therefore able to provide better constraints on $\lambda$ in these problematic cases. Directly modelling the separate components of the CCF provides several separate constraints on the parameters involved in the model, and the geometric calculation of the position of the planet's shadow on the stellar disc helps to remove ambiguity regarding $\lambda$. These two factors lift the degeneracy experienced with the traditional method.

\begin{figure*}
	\subfloat{
		\includegraphics[width=0.48\textwidth]{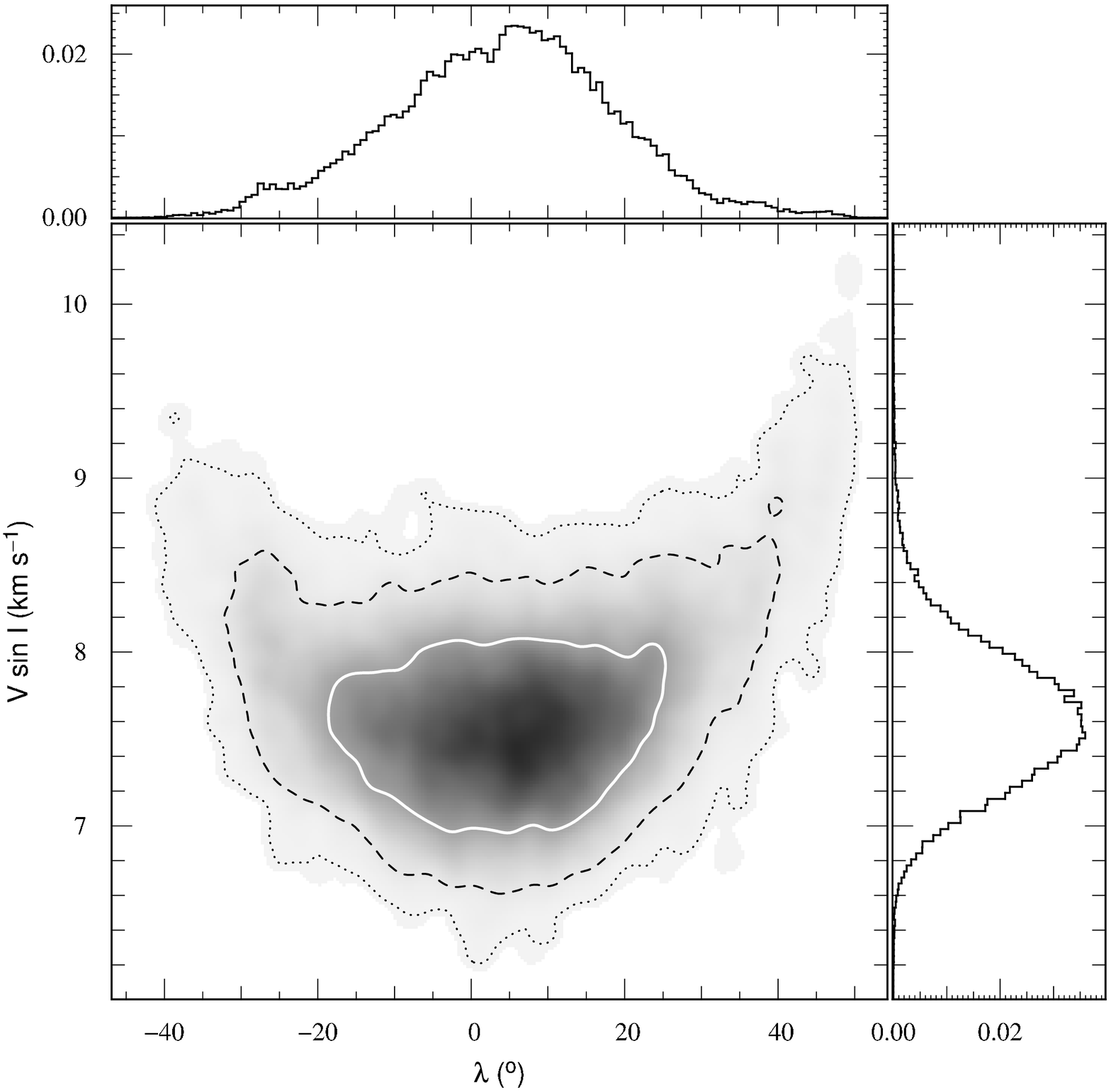}
%		\plotone{fig8a.eps}
		\label{fig:RVdist}}
	\subfloat{
		\includegraphics[width=0.48\textwidth]{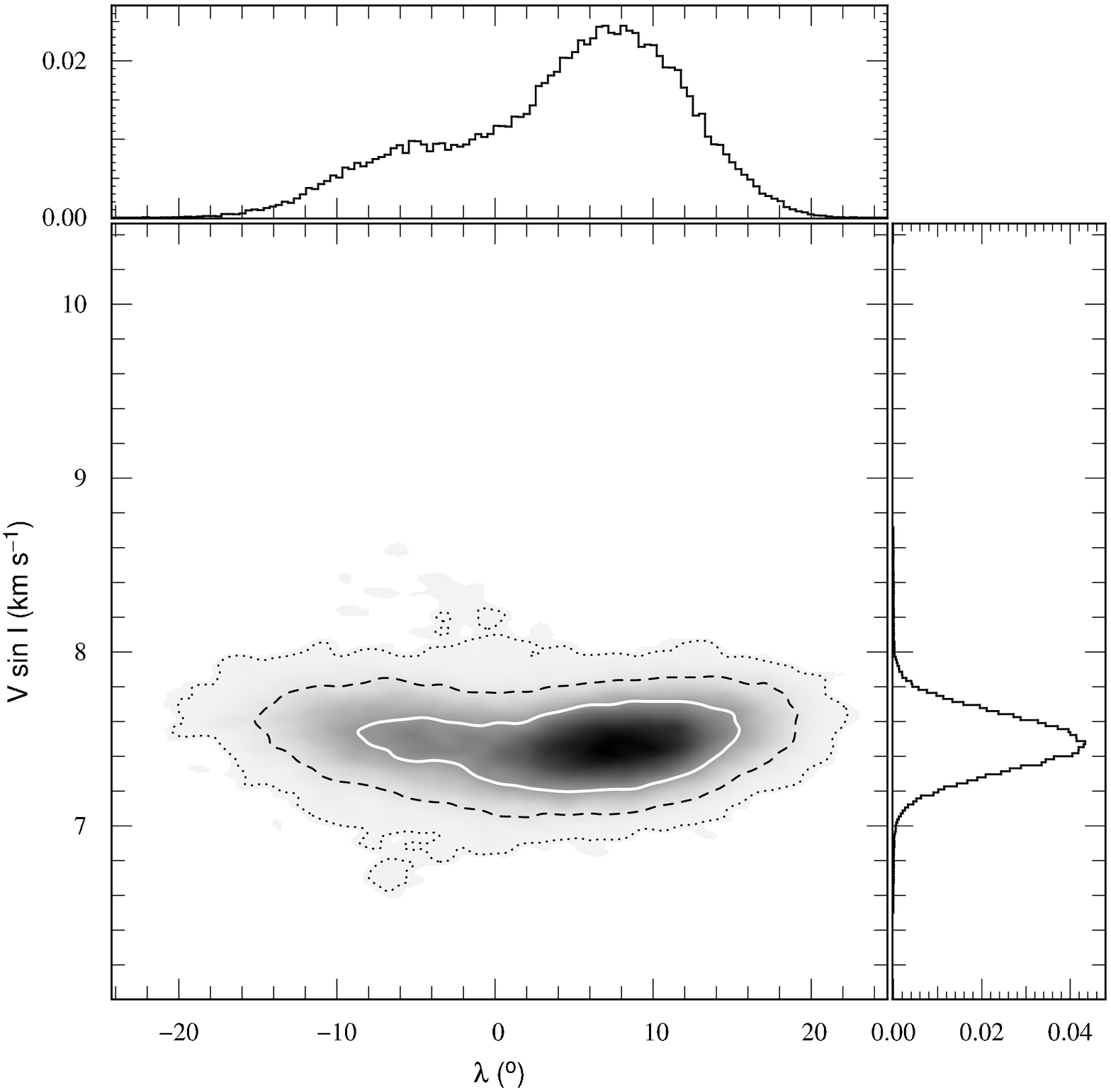}
%		\plotone{fig8b.eps}
		\label{fig:DTdist}}
	\caption{Posterior probability distributions for $v\sin I$ and $\lambda$ for both of the analysis methods discussed in this work. These distributions are for the analysis of the WASP-38 system discussed in Section\,\ref{sec:W38res}. Key as for Fig.\,\ref{fig:W40lam-v}. \textit{Left: }Radial velocity measurement based Rossiter-McLaughlin analysis. \textit{Right: } Doppler Tomography. The difference between the two methods is stark, with the tomographic analysis yielding a much reduced correlation between the parameters.}
	\label{fig:compare}
\end{figure*}

WASP-38, as an example of a system with low impact parameter, provides a reasonable example of the advantages that the tomographic analysis method holds over the standard radial velocity method. Table\,\ref{tab:W38doppres} clearly shows that the error bars on $\lambda$ have been decreased by the use of Doppler tomography, and Fig.\,\ref{fig:compare} shows the change in the relationship between the values of $v\sin I$ and $\lambda$ for the two analysis methods. The two posterior probability density plots show completely different distributions, with that for the radial velocity method showing a clear correlation between the two parameters, with obvious degeneracies in the fitted values. The tomographic distribution, on the other hand, shows very little in the way of correlation, and although there is still some spread in the $\lambda$ distribution the range of $v\sin I$ values has quite clearly been heavily restricted.
 
\section{Conclusions}
\label{sec:conclude}
We have presented measurements of the sky-projected spin-orbit alignment angle for the hot Jupiters WASP-32, WASP-38, and HAT-P-27WASP-40, using both the Rossiter-McLaughlin effect and Doppler tomography. We find that WASP-32 exhibits an alignment angle of $\lambda=10.5^{\circ\,+6.4}_{\,\,\,-5.9}$ (from Doppler tomography) and a rotation speed of $v\sin I=3.9^{+0.4}_{-0.5}$\,km\,s$^{-1}$, indicating an aligned system. The results from our two analysis methods are consistent, and show good agreement, where applicable, with the original discovery paper. For HAT-P-27/WASP-40 we find a much lower rotation speed than suggested by the discovery paper and spectroscopic analysis, $v\sin I=0.6^{+0.7}_{-0.4}$\,km\,s$^{-1}$, but our poor signal-to-noise data allows us to place only weak constraints on the alignment angle. We find $\lambda=24.2^{\circ\,+76.0}_{\,\,\,-44.5}$, which we classify as a non-detection, and are unable to apply the tomography method to the system. For WASP-38 we improve on the previous analysis of \citet{simpson2011}, reducing the uncertainty in $\lambda$ by an order of magnitude, and obtaining $\lambda=7.5^{\circ\,+4.7}_{\,\,\,-6.1}$ and $v\sin I=7.5^{+0.1}_{-0.2}$\,km\,s$^{-1}$ through tomographic analysis. Our results again agree well between the two analysis methods.

We consider the effect of our new results on the complete sample of hot Jupiters for which $\lambda$ has been measured, finding that they either provide support for, or no evidence in opposition to, previously existing trends within the ensemble. We also consider the benefits of using Doppler tomography over the Rossiter-McLaughlin analysis method, and comment that its use is helpful in lifting degeneracies in the fitted solution that arise when using the more traditional method.

\section*{Nota Bene}
We have used the UTC time standard and Barycentric Julian Dates in our analysis. Our results are based on the equatorial solar and jovian radii, and masses taken from AllenÕs Astrophysical Quantities.

\acknowledgments

The WASP Consortium consists of representatives from the Universities of Keele, Leicester, The Open University, Queens University Belfast and St Andrews, along with the Isaac Newton Group (La Palma) and the Instituto de Astrofisica de Canarias (Tenerife). The SuperWASP and WASP-S cameras are operated with funds made available from Consortium Universities and the STFC. TRAPPIST is funded by the Belgian Fund for Scientific Research (Fond National de la Recherche Scientifique, FNRS) under the grant FRFC 2.5.594.09.F, with the participation of the Swiss National Science Foundation (SNF). R.~F~D is supported by CNES. M.~G is an FNRS Research Associate, and acknowledges support from the Belgian Science Policy Office in the form of a Return Grant. I.~B. acknowledges the support of the European Research Council/ European Community under the FP7 through a Starting Grant, as well from Funda\c{c}\~ao para a Ci\^encia e a Tecnologia (FCT), Portugal, through SFRH/BPD/81084/2011 and the project PTDC/CTE-AST/098528/2008. This research has made use of NASA's Astrophysics Data System Bibliographic Services, the ArXiv preprint service hosted by Cornell University, and RenŽ Heller's Holt-Rossiter-McLaughlin Encyclopaedia (www.aip.de/People/RHeller).

\clearpage



\appendix

\section{Journal of observations}
\label{sec:journal}

\begin{table*}
	\caption{RV data for WASP-32 obtained using HARPS. \label{tab:W32harps}}
	\begin{tabular}{lll}
		\tableline \tableline\\
					  & RV 		  & $\sigma_{RV}$ \\ [2pt]
		BJD$_{\rm UTC}$(-2450000) & (km\,s$^{-1}$)& (km\,s$^{-1}$) \\ [2pt]
		\tableline \\
		$5828.581434$ & $17.82469$ & $0.00919$ \\ [2pt]
		$5828.829629$ & $17.83582$ & $0.00647$ \\ [2pt]
		$5829.586912$ & $18.52797$ & $0.00579$ \\ [2pt]
		$5829.807179$ & $18.70852$ & $0.00529$ \\ [2pt]
		$5830.593731$ & $18.43669$ & $0.00780$ \\ [2pt]
		$5830.602145$ & $18.41150$ & $0.00836$ \\ [2pt]
		$5830.611046$ & $18.41026$ & $0.01001$ \\ [2pt]
		$5830.618766$ & $18.38721$ & $0.00942$ \\ [2pt]
		$5830.625803$ & $18.40267$ & $0.00901$ \\ [2pt]
		$5830.633338$ & $18.38290$ & $0.00940$ \\ [2pt]
		$5830.640444$ & $18.38975$ & $0.00933$ \\ [2pt]
		$5830.648303$ & $18.37346$ & $0.00814$ \\ [2pt]
		$5830.656497$ & $18.35622$ & $0.00822$ \\ [2pt]
		$5830.664356$ & $18.36270$ & $0.00816$ \\ [2pt]
		$5830.672365$ & $18.35400$ & $0.00908$ \\ [2pt]
		$5830.680433$ & $18.33837$ & $0.01021$ \\ [2pt]
		$5830.688639$ & $18.34691$ & $0.00943$ \\ [2pt]
		$5830.696567$ & $18.34257$ & $0.00950$ \\ [2pt]
		$5830.704912$ & $18.30751$ & $0.00858$ \\ [2pt]
		$5830.712840$ & $18.30229$ & $0.00884$ \\ [2pt]
		$5830.720907$ & $18.29955$ & $0.00890$ \\ [2pt]
		$5830.729183$ & $18.27168$ & $0.00844$ \\ [2pt]
		$5830.736960$ & $18.24540$ & $0.00829$ \\ [2pt]
		$5830.745178$ & $18.23510$ & $0.00865$ \\ [2pt]
		$5830.753176$ & $18.22897$ & $0.00807$ \\ [2pt]
		$5830.761231$ & $18.22650$ & $0.00802$ \\ [2pt]
		$5830.769298$ & $18.23197$ & $0.00804$ \\ [2pt]
		$5830.777296$ & $18.22548$ & $0.00830$ \\ [2pt]
		$5830.785363$ & $18.21815$ & $0.00815$ \\ [2pt]
		$5830.793546$ & $18.20186$ & $0.00854$ \\ [2pt]
		$5830.801613$ & $18.21736$ & $0.00877$ \\ [2pt]
		$5830.809623$ & $18.19578$ & $0.00891$ \\ [2pt]
		$5830.817771$ & $18.18887$ & $0.00960$ \\ [2pt]
		$5830.825780$ & $18.16949$ & $0.01005$ \\ [2pt]
		$5831.590826$ & $17.85109$ & $0.00602$ \\ [2pt]
		$5831.811104$ & $18.01057$ & $0.00551$ \\ [2pt]
		\tableline \\
	\end{tabular}
\end{table*}

\begin{table*}
	\caption{Out-of-transit RV data for WASP-38 obtained using HARPS. \label{tab:W38harps_oot}}
	\begin{tabular}{lll}
		\tableline \tableline \\
					  & RV 		  & $\sigma_{RV}$ \\ [2pt]
		BJD$_{\rm UTC}$(-2450000) & (km\,s$^{-1}$)& (km\,s$^{-1}$) \\ [2pt]
		\tableline \\
		$5656.783091$ & $-9.51508$ & $0.00337$ \\ [2pt]
		$5657.783221$ & $-9.52550$ & $0.00299$ \\ [2pt]
		$5660.811940$ & $-9.98361$ & $0.00319$ \\ [2pt]
		$5662.834946$ & $-9.64728$ & $0.00423$ \\ [2pt]
		$5680.716630$ & $-9.95563$ & $0.00389$ \\ [2pt]
		$5681.710237$ & $-9.97538$ & $0.00295$ \\ [2pt]
		$5683.728896$ & $-9.59883$ & $0.00285$ \\ [2pt]
		$5714.660893$ & $-9.89326$ & $0.00588$ \\ [2pt]
		$5716.602469$ & $-9.90687$ & $0.00307$ \\ [2pt]
		$5749.637217$ & $-9.96974$ & $0.00642$ \\ [2pt]
		$5753.648280$ & $-9.50884$ & $0.00481$ \\ [2pt]
		$5802.476558$ & $-9.58648$ & $0.00600$ \\ [2pt]
		$5806.489798$ & $-9.80177$ & $0.00370$ \\ [2pt]
		$5809.496470$ & $-9.61798$ & $0.00300$ \\ [2pt]
		\tableline \\
	\end{tabular}
\end{table*}

\begin{table}
	\caption{In-transit RV data for WASP-38 obtained using HARPS. \label{tab:W38harps}}
	\tiny
	\begin{tabular}{lll}
		\hline \\
					  & RV 		  & $\sigma_{RV}$ \\ [2pt]
		BJD$_{\rm UTC}$(-2450000) & (km\,s$^{-1}$)& (km\,s$^{-1}$) \\ [2pt]
		\hline \\
		$5727.508875$ & $-9.72505$ & $0.00658$ \\ [2pt]
		$5727.519373$ & $-9.72057$ & $0.00586$ \\ [2pt]
		$5727.523134$ & $-9.71142$ & $0.00564$ \\ [2pt]
		$5727.527081$ & $-9.70382$ & $0.00585$ \\ [2pt]
		$5727.530912$ & $-9.69675$ & $0.00523$ \\ [2pt]
		$5727.534824$ & $-9.69758$ & $0.00528$ \\ [2pt]
		$5727.538562$ & $-9.69301$ & $0.00529$ \\ [2pt]
		$5727.542879$ & $-9.69153$ & $0.00527$ \\ [2pt]
		$5727.546791$ & $-9.69410$ & $0.00599$ \\ [2pt]
		$5727.550656$ & $-9.68355$ & $0.00613$ \\ [2pt]
		$5727.554603$ & $-9.68966$ & $0.00672$ \\ [2pt]
		$5727.558272$ & $-9.67696$ & $0.00592$ \\ [2pt]
		$5727.562045$ & $-9.69909$ & $0.00646$ \\ [2pt]
		$5727.566397$ & $-9.69542$ & $0.00694$ \\ [2pt]
		$5727.570343$ & $-9.68220$ & $0.00618$ \\ [2pt]
		$5727.574140$ & $-9.70082$ & $0.00615$ \\ [2pt]
		$5727.578017$ & $-9.69531$ & $0.00609$ \\ [2pt]
		$5727.581882$ & $-9.70552$ & $0.00618$ \\ [2pt]
		$5727.585586$ & $-9.70640$ & $0.00618$ \\ [2pt]
		$5727.592484$ & $-9.71415$ & $0.00667$ \\ [2pt]
		$5727.596430$ & $-9.72199$ & $0.00718$ \\ [2pt]
		$5727.600203$ & $-9.71776$ & $0.00651$ \\ [2pt]
		$5727.604046$ & $-9.73514$ & $0.00681$ \\ [2pt]
		$5727.607923$ & $-9.73209$ & $0.00676$ \\ [2pt]
		$5727.611800$ & $-9.73769$ & $0.00745$ \\ [2pt]
		$5727.619335$ & $-9.74676$ & $0.00739$ \\ [2pt]
		$5727.623177$ & $-9.75379$ & $0.00724$ \\ [2pt]
		$5727.627020$ & $-9.76161$ & $0.00671$ \\ [2pt]
		$5727.630746$ & $-9.76295$ & $0.00678$ \\ [2pt]
		$5727.634589$ & $-9.76762$ & $0.00661$ \\ [2pt]
		$5727.638466$ & $-9.76457$ & $0.00690$ \\ [2pt]
		$5727.642667$ & $-9.78228$ & $0.00754$ \\ [2pt]
		$5727.646510$ & $-9.77732$ & $0.00655$ \\ [2pt]
		$5727.650422$ & $-9.78573$ & $0.00707$ \\ [2pt]
		$5727.654183$ & $-9.78749$ & $0.00695$ \\ [2pt]
		$5727.658164$ & $-9.78420$ & $0.00719$ \\ [2pt]
		$5727.661868$ & $-9.79843$ & $0.00745$ \\ [2pt]
		$5727.666231$ & $-9.80281$ & $0.00769$ \\ [2pt]
		$5727.670143$ & $-9.79479$ & $0.00621$ \\ [2pt]
		$5727.673905$ & $-9.79231$ & $0.00634$ \\ [2pt]
		$5727.677747$ & $-9.79598$ & $0.00576$ \\ [2pt]
		$5727.681543$ & $-9.79677$ & $0.00532$ \\ [2pt]
		$5727.685386$ & $-9.80741$ & $0.00569$ \\ [2pt]
		$5727.689656$ & $-9.80222$ & $0.00564$ \\ [2pt]
		$5727.693430$ & $-9.79689$ & $0.00612$ \\ [2pt]
		$5727.697226$ & $-9.79349$ & $0.00704$ \\ [2pt]
		$5727.701242$ & $-9.78573$ & $0.00693$ \\ [2pt]
		$5727.705049$ & $-9.78707$ & $0.00585$ \\ [2pt]
		$5727.708915$ & $-9.78243$ & $0.00568$ \\ [2pt]
		$5727.713197$ & $-9.76337$ & $0.00552$ \\ [2pt]
		$5727.716936$ & $-9.77080$ & $0.00593$ \\ [2pt]
		$5727.720871$ & $-9.77108$ & $0.00529$ \\ [2pt]
		$5727.724713$ & $-9.76885$ & $0.00528$ \\ [2pt]
		$5727.728544$ & $-9.76841$ & $0.00528$ \\ [2pt]
		$5727.732306$ & $-9.76986$ & $0.00514$ \\ [2pt]
		$5727.736484$ & $-9.76921$ & $0.00541$ \\ [2pt]
		$5727.740141$ & $-9.76496$ & $0.00641$ \\ [2pt]
		$5727.744192$ & $-9.76324$ & $0.00720$ \\ [2pt]
		$5727.747999$ & $-9.77120$ & $0.00705$ \\ [2pt]
		$5727.751981$ & $-9.76748$ & $0.00679$ \\ [2pt]
		$5727.755823$ & $-9.76974$ & $0.00648$ \\ [2pt]
		$5727.760256$ & $-9.78057$ & $0.00585$ \\ [2pt]
		$5727.764376$ & $-9.78068$ & $0.00588$ \\ [2pt]
		\hline \\
	\end{tabular}
	\normalsize
\end{table}

\begin{table*}
	\caption{RV data for WASP-40 obtained using HARPS. \label{tab:W40harps}}
	\begin{tabular}{lll}
		\tableline \tableline \\
					  & RV 		  & $\sigma_{RV}$ \\ [2pt]
		BJD$_{\rm UTC}$(-2450000) & (km\,s$^{-1}$)& (km\,s$^{-1}$) \\ [2pt]
		\tableline \\
		$5686.691040$ & $-15.68740$ & $0.00531$ \\ [2pt]
		$5691.594153$ & $-15.83574$ & $0.00593$ \\ [2pt]
		$5691.621178$ & $-15.84591$ & $0.00639$ \\ [2pt]
		$5691.784798$ & $-15.82559$ & $0.00548$ \\ [2pt]
		$5692.677700$ & $-15.67883$ & $0.00455$ \\ [2pt]
		$5692.747097$ & $-15.67360$ & $0.00411$ \\ [2pt]
		$5693.571851$ & $-15.75439$ & $0.00814$ \\ [2pt]
		$5693.577950$ & $-15.75479$ & $0.00878$ \\ [2pt]
		$5693.584107$ & $-15.74939$ & $0.00838$ \\ [2pt]
		$5693.590207$ & $-15.75567$ & $0.00843$ \\ [2pt]
		$5693.596480$ & $-15.74962$ & $0.00807$ \\ [2pt]
		$5693.602579$ & $-15.75667$ & $0.00803$ \\ [2pt]
		$5693.608737$ & $-15.76548$ & $0.00767$ \\ [2pt]
		$5693.614952$ & $-15.75844$ & $0.00817$ \\ [2pt]
		$5693.621167$ & $-15.76064$ & $0.00735$ \\ [2pt]
		$5693.627093$ & $-15.75956$ & $0.00822$ \\ [2pt]
		$5693.633296$ & $-15.77570$ & $0.00901$ \\ [2pt]
		$5693.639569$ & $-15.76272$ & $0.00858$ \\ [2pt]
		$5693.645669$ & $-15.77900$ & $0.00757$ \\ [2pt]
		$5693.651884$ & $-15.77096$ & $0.00700$ \\ [2pt]
		$5693.657925$ & $-15.77981$ & $0.00764$ \\ [2pt]
		$5693.664152$ & $-15.76408$ & $0.00737$ \\ [2pt]
		$5693.670263$ & $-15.76868$ & $0.00751$ \\ [2pt]
		$5693.676467$ & $-15.76874$ & $0.00774$ \\ [2pt]
		$5693.682682$ & $-15.76226$ & $0.00709$ \\ [2pt]
		$5693.688781$ & $-15.78715$ & $0.00662$ \\ [2pt]
		$5693.694927$ & $-15.77673$ & $0.00671$ \\ [2pt]
		$5693.701142$ & $-15.78574$ & $0.00677$ \\ [2pt]
		$5693.707184$ & $-15.78466$ & $0.00663$ \\ [2pt]
		$5693.713387$ & $-15.76800$ & $0.00666$ \\ [2pt]
		$5693.719603$ & $-15.78821$ & $0.00713$ \\ [2pt]
		$5693.725633$ & $-15.77244$ & $0.00792$ \\ [2pt]
		$5693.787541$ & $-15.78723$ & $0.00474$ \\ [2pt]
		$5694.567017$ & $-15.85210$ & $0.00448$ \\ [2pt]
		$5695.576799$ & $-15.69093$ & $0.00362$ \\ [2pt]
		$5695.757478$ & $-15.67832$ & $0.00406$ \\ [2pt]
		\tableline \\
	\end{tabular}
\end{table*}

\end{document}